\documentclass[%
 reprint,
 superscriptaddress,
 amsmath,amssymb,
 aps,
 prl,
]{revtex4-2}

\usepackage{graphicx}
\usepackage{dcolumn}
\usepackage{bm}
\usepackage{hyperref}
\usepackage{url}
\hypersetup{hidelinks}
\usepackage[dvipsnames]{xcolor}
\usepackage{makecell}
\usepackage{physics}
\usepackage{upgreek}
\usepackage{float}
\usepackage{amsmath}
\usepackage{siunitx}
\begin{document}

\title{Highly spin-polarized carriers and strong ferromagnetism \\  in doped perovskite antiferromagnetic semiconductors}

\author{Hong Jian Zhao}
 \affiliation{Key Laboratory of Material Simulation Methods and Software of Ministry of Education, College of Physics, Jilin University, Changchun 130012, China}
 \affiliation{Key Laboratory of Physics and Technology for Advanced Batteries (Ministry of Education), College of Physics, Jilin University, Changchun 130012, China}
 \affiliation{State Key Laboratory of Superhard Materials, College of Physics, Jilin University, Changchun 130012, China}
 \affiliation{International Center of Future Science, Jilin University, Changchun 130012, China}
 \author{Longju Yu}
 \affiliation{Key Laboratory of Material Simulation Methods and Software of Ministry of Education, College of Physics, Jilin University, Changchun 130012, China}
 \author{Yanchao Wang}
 \email{wyc@calypso.cn}
 \affiliation{Key Laboratory of Material Simulation Methods and Software of Ministry of Education, College of Physics, Jilin University, Changchun 130012, China}
 \affiliation{State Key Laboratory of Superhard Materials, College of Physics, Jilin University, Changchun 130012, China}
\author{Laurent Bellaiche}
\affiliation{Physics Department and Institute for Nanoscience and Engineering, University of Arkansas, Fayetteville, Arkansas 72701, USA}
\author{Yanming Ma}
\email{mym@jlu.edu.cn}
 \affiliation{Key Laboratory of Material Simulation Methods and Software of Ministry of Education, College of Physics, Jilin University, Changchun 130012, China}
 \affiliation{International Center of Future Science, Jilin University, Changchun 130012, China}
 \affiliation{State Key Laboratory of Superhard Materials, College of Physics, Jilin University, Changchun 130012, China}

\date{\today}

\begin{abstract}
In semiconductor spintronics, the generation of highly spin-polarized carriers and the efficient probe of spin order (due to strong ferromagnetism) -- at or above room temperature -- are crucial because it allows for the design of spin-based semiconductor devices. Usually, such goals were fulfilled in room-temperature ferromagnetic semiconductors, being rare materials in nature. While room-temperature antiferromagnetic semiconductors are plentiful, the possibility for creating highly spin-polarized carriers and strong ferromagnetism in these materials remain to be unraveled. Here, we explore such a possibility by first-principles simulations, working with CaTcO$_3$ and NaOsO$_3$ perovskites -- being room-temperature antiferromagnetic semiconductors. We find that doping them by electrons or holes results in these materials to be highly spin-polarized, carrying enormous ferromagnetic moments. Doping electrons with moderate carrier density can yield strong ferromagnetism in them, with the ferromagnetic moments being comparable to that in typical ferromagnetic semiconductors. Our work thus indicates the merit of perovskite antiferromagnetic semiconductors in spintronics -- for a possible replacement of ferromagnetic semiconductors.
\end{abstract}

\maketitle

\noindent
\textit{Introduction.--} Ferromagnetic  semiconductors are of vital significance in semiconductor spintronics~\cite{science2001,spintronic2,natphot2021,afmnsr}. Materials of this kind present giant Zeeman-type spin splittings and strong ferromagnetism, where the former allow for the generation of highly spin-polarized carriers and the latter permits the efficient probe of spin order~\cite{science2001,spintronic2,afm2018}. One shortcoming of ferromagnetic semiconductors is that the Curie temperatures of these materials are usually far lower than room temperature~\cite{science2001,ferri2021}, making the ferromagnetic semiconductors lose their practical applications. In this regard, ferrimagnetic semiconductors and related materials -- such as asymmetric and bipolar antiferromagnetic semiconductors~\footnote{The asymmetric antiferromagnetic semiconductors and bipolar antiferromagnetic semiconductors reported in Refs.~\cite{wang2022,yang2015}, being ferrimagnetic semiconductors in essence, involve more than one magnetic sublattice with almost cancelled magnetic moments. The densities of states of these materials (together with ferrimagnetic semiconductors) resemble those in ferromagnets, with giant splittings between spin-up and spin-down channels. Consequently, the doped carriers in these materials can be highly spin-polarized, significantly tuning the magnetic moments.}, being still rare at room temperature -- may come to the rescue, serving as the ``weak version'' of ferromagnets (see, {\it e.g.}, Refs.~\cite{ferri2021,wang2022,ferri2008,yang2015,afmnsr}). 
On the other hand, there are a variety of perovskite semiconductors being antiferromagnetic with N\'eel temperatures being above room temperature~\cite{noncollinear,afmrmp2018}. Some of these perovskites ({\it e.g.}, YFeO$_3$ and CaTcO$_3$) can host Zeeman-type spin splittings and weak ferromagnetism~\cite{zhou2020,zeemanafm}. Typically, the Zeeman-type spin splitting can reach $\sim$78 meV in CaTcO$_3$ perovskite~\cite{zeemanafm}, being sizable -- albeit a tiny value compared with $\sim$2 eV in YTiO$_3$~\cite{liu2022}; The magnitudes of magnetic moments in antiferromagnets are much smaller than those in ferromagnets ({\it e.g.}, $\sim$0.06 $\mu_B/f.u.$ in antiferromagnetic YFeO$_3$~\cite{zhou2020} {\it versus} $\sim$0.8 $\mu_B/f.u.$ in ferromagnetic YTiO$_3$~\cite{ytio3,ytio32}). At all events, this suggests a route to utilize perovskite antiferromagnetic semiconductors as alternatives for ferromagnetic semiconductors in spintronics, if the following questions are positively answered: (i) Is it possible to generate highly spin-polarized carriers in antiferromagnetic semiconductors? (ii) Is there an avenue to realize strong ferromagnetism in antiferromagnetic semiconductors?

In this Letter, we address the aforementioned questions by first-principles simulations. We focus on perovskite CaTcO$_3$ and NaOsO$_3$, whose N\'eel temperatures are $\sim$800 K~\cite{catco3} and $\sim$410 K~\cite{naoso3,naoso32,naoso33,naoso3mit}, respectively. Below N\'eel temperatures, these materials are all antiferromagnetic semiconductors~\cite{catco3prb,ctou2,naoso3,naoso32,naoso33,naoso3mit}. 
We show that  doping  CaTcO$_3$ and NaOsO$_3$ by electrons or holes leads them to be highly spin-polarized, due to the sizable Zeeman-type spin splittings of energy levels around their valence band maximum (VBM) and conduction band minimum (CBM). The doped carriers, carrying enormous ferromagnetic moments, significantly enhance the ferromagnetism in antiferromagnetic CaTcO$_3$ and NaOsO$_3$. In particular, doping electrons with the carrier density of 0.125 $e/f.u.$ ($e$ being the charge of electron) gives rise to ferromagnetic moments of $\sim$0.3 $\mu_B/f.u.$ in CaTcO$_3$ and $\sim$0.2 $\mu_B/f.u.$ in NaOsO$_3$, an order of magnitude being comparable to those in perovskite ferromagnetic semiconductors ({\it e.g.}, $\sim$0.8 $\mu_B/f.u.$ in YTiO$_3$~\cite{ytio3,ytio32}).

\begin{figure}[htbp]
\centering
\includegraphics[width=1.0\linewidth]{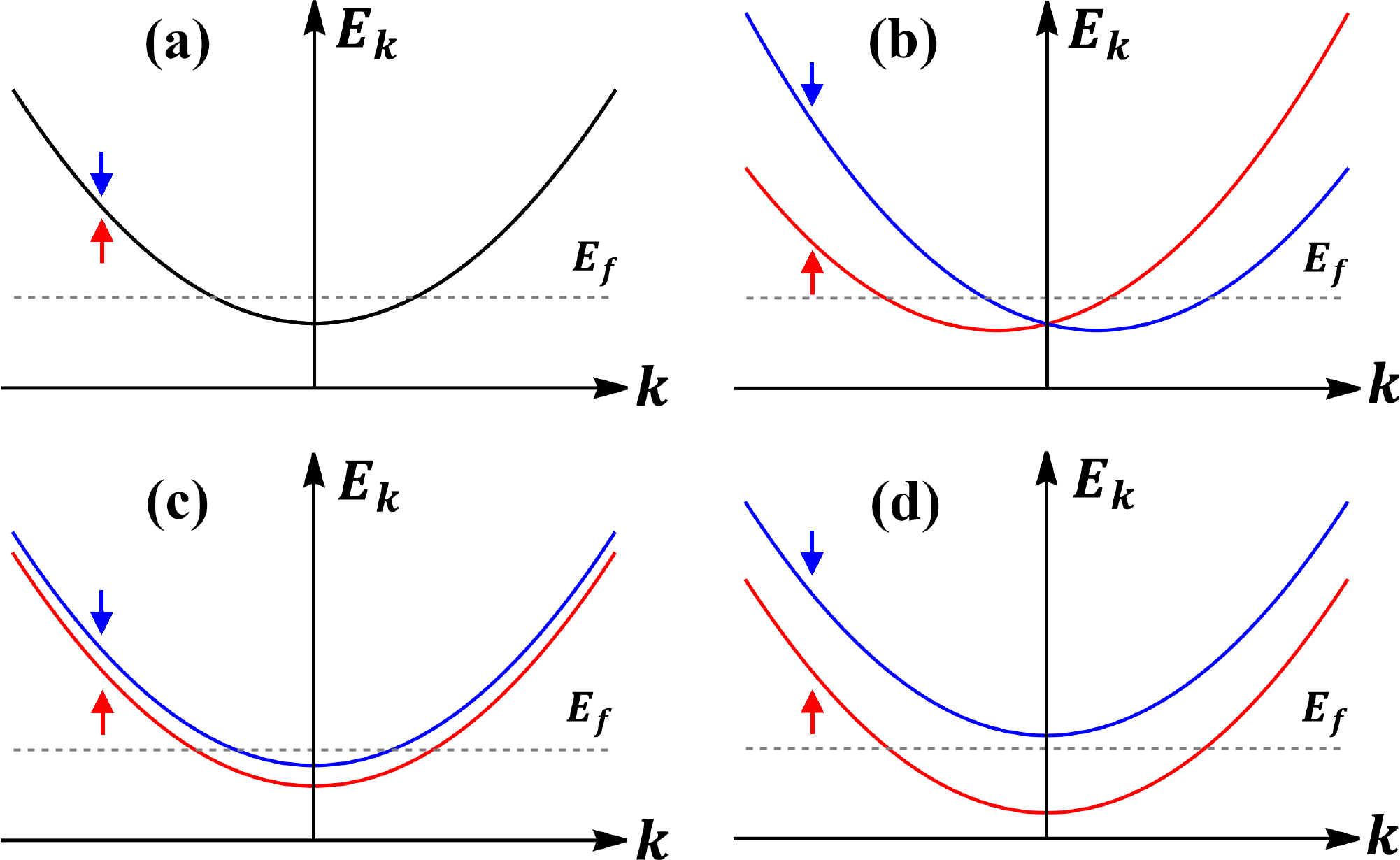}
\caption{\label{fig:Zeemansketch}
Schematizations of the band structures ({\it i.e.}, band energies $E_{\boldsymbol{k}}$ versus the wave vector $\boldsymbol{k}$) in electron-doped semiconductors with null spin splitting [panel (a)], Rashba-/Dresselhaus-type spin splitting [panel (b)], tiny Zeeman-type spin splitting [panel (c)], and sizable Zeeman-type spin splitting [panel (d)]. The split energy levels with positive and negative spin magnetization values are sketched in red and blue curves (and marked by red and blue arrows), respectively. The horizontal dash line in each panel denotes the Fermi energy level $E_f$.}
\end{figure}

\noindent
\textit{Spin splittings \textit{versus} spin-polarized carriers.--} To begin with, we revisit the possibilities for creating spin-polarized carriers in semiconductors by doping. In centrosymmetric non-magnetic semiconductors, the electron's spin energy levels associated with $+S_\alpha$ and $-S_\alpha$ ($\alpha=x,y,z$), the spin magnetization of the electron, are degenerate everywhere in $\boldsymbol{k}$ space ($\boldsymbol{k}$ being the wave vector). Doping such semiconductors is not expected to yield spin-polarized carriers because the carriers occupy $+S_\alpha$ and $-S_\alpha$ states in a symmetrical manner [see Fig.~\ref{fig:Zeemansketch}(a)]. As for the non-centrosymmetric non-magnetic semiconductors, a $\boldsymbol{k}$-dependent effective magnetic field $\mathbf{B}^\mathrm{eff}(\boldsymbol{k})$ is emerged because of the spin-orbit coupling (SOC)~\cite{spinsplitting,spinsplitting2}. Such a field couples with electron's spin via $\mathbf{B}^\mathrm{eff}(\boldsymbol{k})\cdot\boldsymbol{\sigma}$~\cite{spinsplitting} and yields, {\it e.g.}, Rashba-/Dresselhaus-type spin splittings~\cite{rashbajetp,dresselhausoriginal}, where $\boldsymbol{\sigma}\equiv(\sigma_x,\sigma_y,\sigma_z)$ is the vector formed by Pauli matrix $\sigma_\alpha$ ($\alpha=x,y,z$). In other words, the $+S_\alpha$ and $-S_\alpha$ levels become non-degenerate at some non-zero $\boldsymbol{k}$ [see Fig.~\ref{fig:Zeemansketch}(b)]. Nonetheless, the lack of magnetism indicates the odd-function feature of $\mathbf{B}^\mathrm{eff}(\boldsymbol{k})$, namely, $\mathbf{B}^\mathrm{eff}(-\boldsymbol{k})=-\mathbf{B}^\mathrm{eff}(\boldsymbol{k})$~\cite{spinsplitting}; As a consequence, each $\pm S_\alpha$ state at any wave vector $\boldsymbol{k}$ can find its degenerate partner ({\it i.e.}, $\mp S_\alpha$ state) at the corresponding $-\boldsymbol{k}$. In such sense, the doped carriers are not spin-polarized either in non-centrosymmetric non-magnetic semiconductors [see Fig.~\ref{fig:Zeemansketch}(b)].

In ferromagnetic or ferrimagnetic materials, there exists an effective magnetic field $B^\mathrm{eff}_\alpha$ being independent of $\boldsymbol{k}$. 
The interaction between $B^\mathrm{eff}_\alpha$ field and electron's spin, indicated by $B^\mathrm{eff}_\alpha \sigma_\alpha$, causes Zeeman-type spin splittings in these materials~\cite{spintronic,spintronic2}. If so, the degeneracy between $+S_\alpha$ and $-S_\alpha$ state ($\alpha$ being a particular direction determined by $B^\mathrm{eff}_\alpha$) can be broken by the $B^\mathrm{eff}_\alpha$ effective magnetic field [see Fig.~\ref{fig:Zeemansketch}(d)]. Recent studies indicate that some antiferromagnetic semiconductors showcase weak ferromagnetism together with 
Zeeman-type spin splittings (see, {\it e.g.}, Ref.~\cite{zeemanafm,zungerspinsplitting,zungerspinsplitting2}). In such sense, the doped carriers in antiferromagnetic semiconductors occupy the $+S_\alpha$ and $-S_\alpha$ states asymmetrically and thus can be spin-polarized~\cite{qi2022,zhao2022zeeman}. Antiferromagnetic semiconductors with tiny Zeeman-type spin splittings around the CBM (VBM)~\footnote{In semiconductors, electron doping or hole doping occurs around the CBM and VBM, respectively.} gain doped electrons (holes) that are slightly spin-polarized [Fig.~\ref{fig:Zeemansketch}(c)], while semiconductors with sizable Zeeman-type splitting splittings yield carriers being significantly (even 100\%) spin-polarized [Fig.~\ref{fig:Zeemansketch}(d)]. Provided that the Zeeman-type spin splittings are large enough around the VBM or CBM, the doped carriers may carry sizable ferromagnetic moment -- reinforcing the weak ferromagnetism in antiferromagnetic semiconductors.

\begin{figure}[htbp]
\centering
\includegraphics[width=1.0\linewidth]{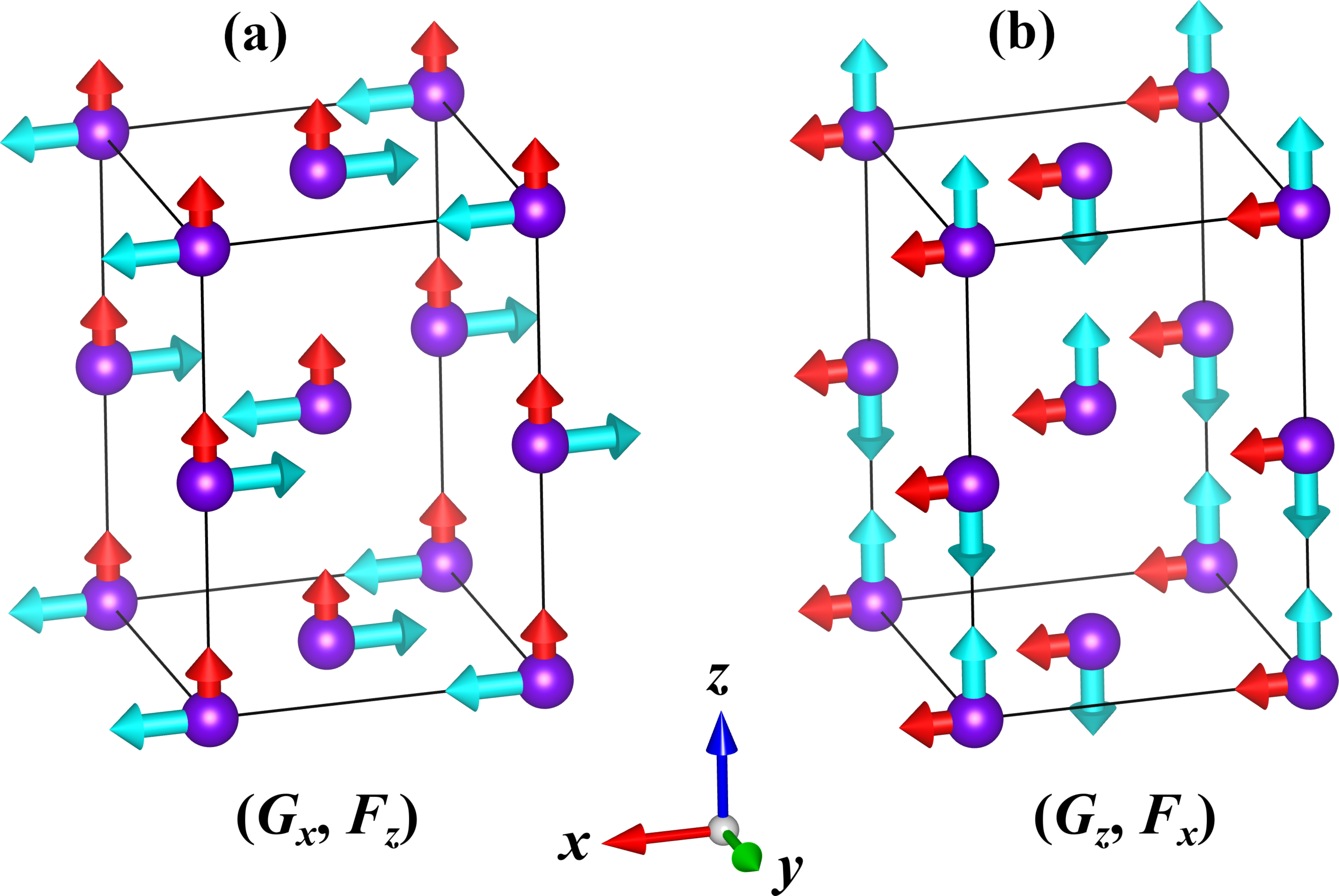}
\caption{\label{fig:ABX3}
Schematizations of the $(G_x, F_z)$ and $(G_z, F_x)$ magnetic configurations in $Pbnm$ $ABX_3$ perovskite. In panels (a) and (b), the predominant $G$-ype and the weak $F$-type vectors (carried by $B$ ions -- represented by purple spheres) are denoted by cyan and red arrows, respectively. For displaying clarify, the $A$ and $X$ ions are not shown in the schematizations. }
\end{figure}

\noindent
\textit{Doped perovskite antiferromagnets: the spin-polarized carriers.--} In the following, we will demonstrate that our aforementioned scenario can be realized in antiferromagnetic $ABX_3$ perovskites. We recall that various $ABX_3$ perovskites ({\it e.g.}, rare-earth orthoferrites~\cite{noncollinear,rfeo3}, rare-earth orthochromates~\cite{noncollinear,rcro3}, SrTcO$_3$~\cite{srtco3}, CaTcO$_3$~\cite{catco3} and NaOsO$_3$~\cite{naoso3,naoso32}) adopt the $Pbnm$ crystallographic space group with the predominant $G$-type antiferromagnetic vectors carried by $B$ ions. Symmetry analyses and experiments~\cite{noncollinear,rfeo3,rcro3,bellaiche2012,zhao2021} indicate that two magnetic configurations -- termed as $(G_x, F_z)$ and $(G_z, F_x)$ magnetic configurations and sketched in Fig.~\ref{fig:ABX3} -- can present weak ferromagnetism (due to spin canting) in $Pbnm$ perovskite with $G$-type antiferromagnetism. To be specific, aligning the predominant $G$-type vectors along $x$ direction (denoted by $G_x$) causes weak ferromagnetic vectors along $z$ (i.e., being $F$-type and denoted by $F_z$). Similarly, the predominant $G_z$ antiferromagnetic vectors yield the weak ferromagnetic $F_x$ vectors. Theories suggest that $(G_x, F_z)$ and $(G_z, F_x)$ magnetic configurations accommodate Zeeman-type spin splittings that are microscopically rooted in SOC, where the magnitudes of Zeeman-type spin splittings depend on the strength of SOC~\cite{zeemanafm}. The split spin energy levels are associated with spin magnetization $\pm S_z$ for $(G_x, F_z)$ configuration and $\pm S_x$ for $(G_z, F_x)$ configuration.

\begin{figure}[htbp]
\centering
\includegraphics[width=1.0\linewidth]{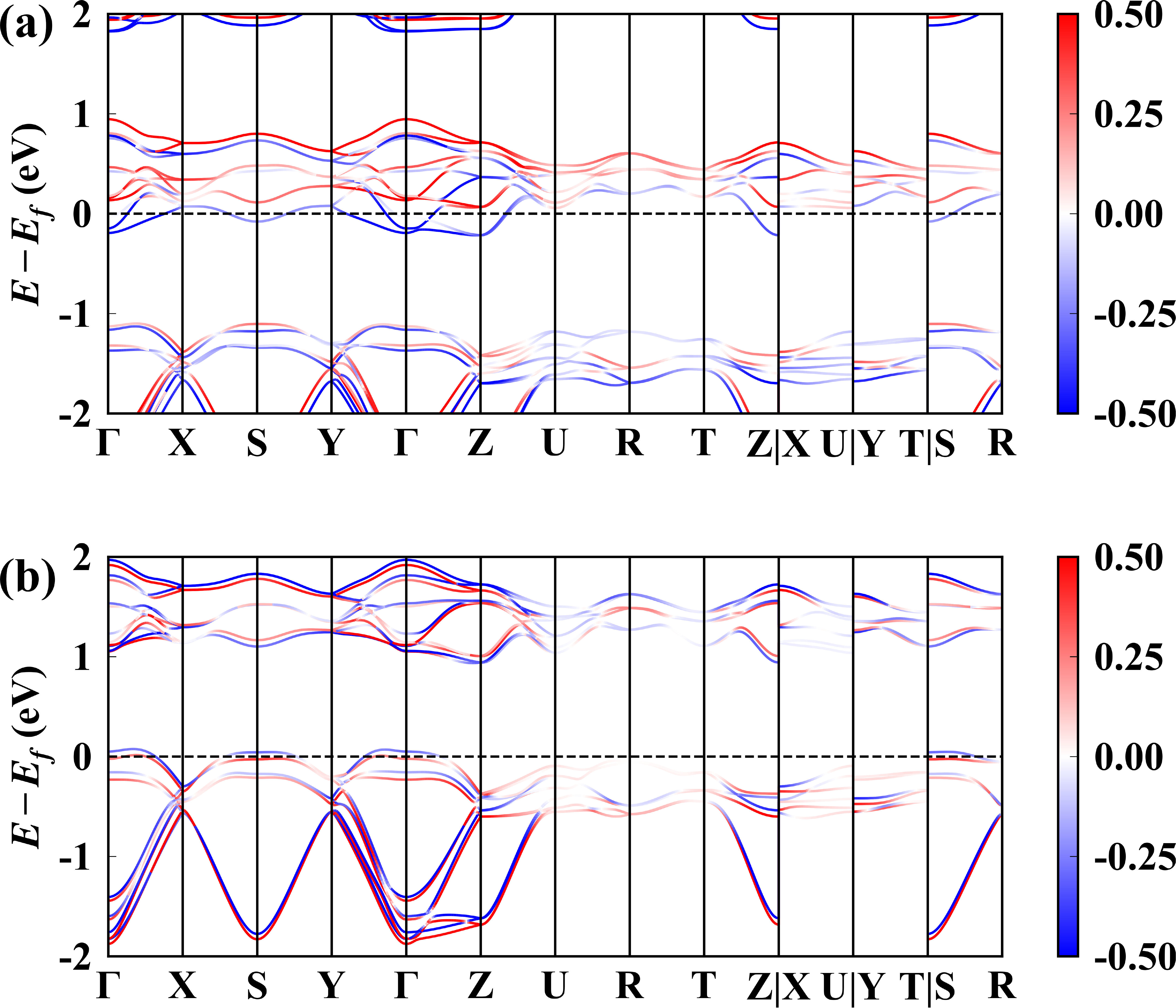}
\caption{\label{fig:catco3gx}
The band structures of CaTcO$_3$ with the $(G_x,F_z)$ magnetic configuration. The zero energy is set as the Fermi level $E_f$. Panels (a) and (b) correspond to electron ($n_{free}=+0.125~e/f.u.$) and hole ($n_{free}=-0.125~e/f.u.$) dopings, respectively. The color bar denotes the spin magnetization $\pm S_z$ (positive value for $+S_z$ and negative value for $-S_z$). In the horizontal axis, the $\boldsymbol{k}$-points in the high-symmetry path are represented by the reduced coordinates, with respect to the reciprocal lattice vectors.}
\end{figure}

Next, we select NaOsO$_3$, CaTcO$_3$ and YFeO$_3$ antiferromagnetic perovskites ($G$-type) as our platforms, where NaOsO$_3$ exhibits strong SOC, CaTcO$_3$ moderate SOC, and YFeO$_3$ weak SOC. We calculate the band structures of these materials with respect to $(G_x,F_z)$ and $(G_z,F_x)$ magnetic configurations, upon carrier doping (carrier density $n_{free}$ being $\pm0.125~e/f.u.$) -- the positive and negative $n_{free}$ values corresponding to electron and hole dopings, respectively. The band structures of CaTcO$_3$ with $(G_x,F_z)$ magnetic configuration are shown in Fig.~\ref{fig:catco3gx}. Sizable Zeeman-type spin splittings occur around the CBM (respectively, VBM) of CaTcO$_3$ doped with electrons (respectively, holes).
Upon electron doping, the extra electrons predominantly occupy the $-S_z$ states with tiny concentration occupying $+S_z$ states [Fig.~\ref{fig:catco3gx}(a)]. As a consequence, the $-S_z$ spin-polarized electrons are injected into the material. The hole doping in CaTcO$_3$ [with $(G_x,F_z)$ configuration], similarly, creates holes that mostly occupy the $-S_z$ states, yielding $-S_z$ spin-polarized holes. Moving to the $(G_z,F_x)$ configuration, the situations regarding Zeeman-type spin splittings are similar to those in $(G_x,F_z)$ configuration, as shown in Fig.~S1 of the Supplementary Material (SM)~\footnote{See Supplementary Material at {\color{blue}a link} for the methods used in the present work, and some numerical results regarding NaOsO$_3$, CaTcO$_3$ and YFeO$_3$.} (Note that our SM contains Refs.~\cite{kresse1996efficient,kresse1999ultrasoft,blochl1994projector,ldaca,ldau,ctojpcm,gfou3,momma2011vesta,seekpath,seekpath2,seekpath3,pyprocar,pyprocar2,hunter2007matplotlib,magndata,magndata2,magndata3,jpcmnoncollinear,exchangej,liupt2020}). In the $(G_z,F_x)$ configuration of CaTcO$_3$, the doped electrons or holes prefer occupying the $+S_x$ states rather than the $-S_x$ states.

Our above discussion is basically valid for NaOsO$_3$ with $(G_x,F_z)$ and $(G_z,F_x)$ magnetic configurations (see Fig.~S2 of the SM). By saying ``\textit{basically}'' we mean that (i) the carriers in NaOsO$_3$ predominantly occupy the $\pm S_\alpha$ level [\textit{e.g.}, $\alpha=z$ in $(G_x,F_z)$ configuration or $\alpha=x$ in $(G_z,F_x)$] -- resembling the cases in CaTcO$_3$, and (ii) a small amount of carriers may also occupy $\mp S_\alpha$ level -- the partner of the $\pm S_\alpha$ level. In such sense, the carriers in NaOsO$_3$ are partially spin-polarized. As for YFeO$_3$, the SOC is negligible and the Zeeman-type spin splittings are tiny. Figure~S3 in the SM indicates that the spin magnetization $\lvert S_\alpha \rvert$ in the vicinity of the doping level [$\alpha=z$ for $(G_x,F_z)$ and $\alpha=x$ for $(G_z,F_x)$] is almost vanishing. Furthermore, the doped carriers occupy $+S_\alpha$ and $-S_\alpha$ states in a nearly symmetrical manner (see Fig.~S4 of the SM). Hence, the spin polarization of the doped carriers in YFeO$_3$ is rather insignificant.

\noindent
\textit{Doped perovskite antiferromagnets: the enhanced ferromagnetism.--} In the undoped $Pbnm$ NaOsO$_3$, CaTcO$_3$ and YFeO$_3$, the predominant $G$-type antiferromagnetic vectors tend to align along the $x$ axis (compared with $y$ and $z$ axes), as predicted by our first-principles calculations. This yields the $(G_x,F_z)$ magnetic configuration in these materials and coincides with the experiments (see \textit{e.g.}, Refs.~\cite{zhou2020,catco3,naoso32} and Section I of the SM). Furthermore, the calculated magnetic moments $(M_z, M_x)$ for NaOsO$_3$, CaTcO$_3$ and YFeO$_3$ are $(0.016,-0.003)$, $(0.057,-0.059)$ and $(0.066,-0.063)$ $\mu_B/f.u.$, respectively, with (i) $M_z$ being associated with $F_z$ in the $(G_x,F_z)$ configuration, and (ii) $M_x$ being associated with $F_x$ in the $(G_z,F_x)$ configuration~\footnote{The $M_x$ and $M_z$ are contributed by the both the spin and orbital magnetic moments.}.

\begin{figure}[htbp]
\centering
\includegraphics[width=1.0\linewidth]{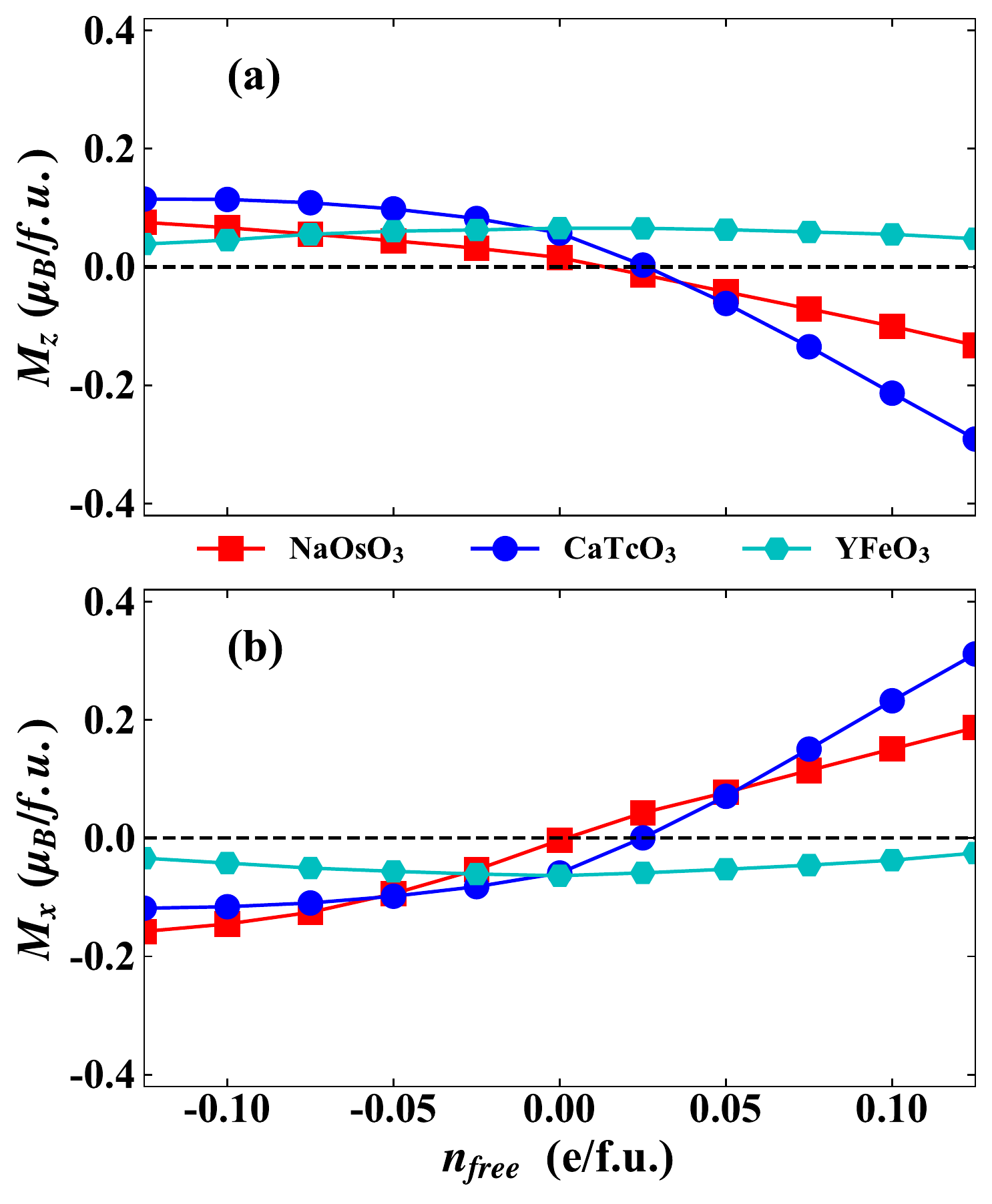}
\caption{\label{fig:dopedABX3}
The magnetic moments of CaTcO$_3$, NaOsO$_3$ and YFeO$_3$ as a function of carrier density $n_{free}$. Panels (a) and (b) correspond to the $(G_x,F_z)$ and $(G_z,F_x)$ magnetic configurations, respectively. }
\end{figure}

We now explore the effect of carrier doping on ferromagnetism in these perovskite antiferromagnets. First of all, our first-principles calculations, at the level of collinear magnetism, show that the $G$-type magnetic phase in NaOsO$_3$, CaTcO$_3$ and YFeO$_3$ are more stable upon doping with $n_{free}=\pm 0.125~e/f.u.$, compared with the $A$-, $C$-, and $F$-type magnetic phases (see Section III of the SM). That NaOsO$_3$ becomes a $G$-type antiferromagnetic metal upon electron doping is basically consistent with the results shown in Refs.~\cite{hubbard2017,naoso3model2}. In the following, our discussion will be limited within the $(G_x,F_z)$ and $(G_z,F_x)$ magnetic configurations.

Figure~\ref{fig:dopedABX3}(a) shows the $M_z$ -- in the $(G_x,F_z)$ magnetic configuration -- of CaTcO$_3$, NaOsO$_3$ and YFeO$_3$ as a function of $n_{free}$. In CaTcO$_3$, doping electrons (holes) reinforces its ferromagnetic component along the $-z$ direction ($+z$ direction). This can be interpreted in the following way. As shown in Fig.~\ref{fig:catco3gx}(a), the doped electrons are $-S_z$ polarized (see above) which carries the ferromagnetic moments along $-z$ direction. As for hole doping, the holes are $-S_z$ polarized as well; Yet, the hole doping is equivalent to removing electrons from the $-S_x$ states of CaTcO$_3$, creating the ferromagnetic moments along $+z$ direction. Regarding the $(G_z,F_x)$ configuration of CaTcO$_3$, the doping effect on ferromagnetism is readily understood by examining Fig.~\ref{fig:dopedABX3}(b) and Fig.~S1 of the SM. Furthermore, the trends for the magnetic moments of NaOsO$_3$ (versus $n_{free}$) follow those of CaTcO$_3$. The physical interpretation is akin to the case of CaTcO$_3$ (see Fig.~S2 of the SM). Unlike CaTcO$_3$ and NaOsO$_3$, the magnetic moments of YFeO$_3$ remain inert upon carrier doping for both $(G_x,F_z)$ and $(G_z,F_x)$ magnetic configurations (see Fig.~\ref{fig:dopedABX3}). This results from the insignificant spin polarization of the doped carriers in YFeO$_3$, and coincides with our discussion in the Section ``\textit{Doped perovskite antiferromagnets: the spin-polarized carriers}''.

To finish this section, we emphasize that carrier doping can significantly enhance the ferromagnetism in antiferromagnetic CaTcO$_3$ and NaOsO$_3$. Strikingly, doping CaTcO$_3$ with electrons ($n_{free}=+0.125~e/f.u.$) yields ferromagnetic moment of the order of $\sim0.3~\mu_B/f.u.$ in both $(G_x,F_z)$ and $(G_z,F_x)$ magnetic configurations. As for NaOsO$_3$, doping electrons with $n_{free}=+0.125~e/f.u.$ results in $M_x$ of $\sim0.2~ \mu_B/f.u.$ in the $(G_z,F_x)$ configuration. Such ferromagnetic moments are comparable to those in the typical ferromagnetic perovskites ({\it e.g.}, $\sim0.8~\mu_B/f.u.$ in YTiO$_3$~\cite{ytio3,ytio32}). For CaTcO$_3$ and NaOsO$_3$, the electron doping level $n_{free}=0.125~e/f.u.$ corresponds to the volume concentrations of $\sim2\times 10^{21}$ cm$^3$, a concentration that could be accessible by electrostatic doping (see, {\it e.g.}, Refs.~\cite{doping1,doping2}). Furthermore, we are aware of an avenue to realizing the electron doping in NaOsO$_3$ ({\it i.e.}, via the replacement of Na with Mg), as proposed by Ref.~\cite{naoso3dop} -- focusing on the insulator-to-metal transition in NaOsO$_3$ driven by doping.

\noindent
\textit{Summary and perspective.--} By first-principles simulations, we have shown that CaTcO$_3$ and NaOsO$_3$ antiferromagnetic perovskites enable the creation of highly spin-polarized carriers by either electron doping or hole doping, thanks to the sizable Zeeman-type spin splittings of the electrons' energy levels in the vicinity of CBM or VBM. The doped electrons or holes carry enormous ferromagnetic moments and significantly enhance the ferromagnetism in antiferromagnetic CaTcO$_3$ and NaOsO$_3$, compared to their undoped neutral cases. Promisingly, the N\'eel temperatures of CaTcO$_3$ and NaOsO$_3$ are all above the room temperature. This implies a route to (i) inject highly spin-polarized carriers (and generate spin current), and (ii) create sizable ferromagnetic moments in doped perovskite antiferromagnetic semiconductors, at room temperature and in the absence of magnetic field. As a perspective, it is possible to foreseen the importance of room-temperature perovskite antiferromagnetic semiconductors -- exhibiting sizable Zeeman-type spin splittings around their VBM and CBM -- in spintronics. We hope that our work can motivate the utilization of such antiferromagnetic semiconductors in the design and fabrication of room-temperature spintronic devices.

\noindent
\textit{Acknowledgements.--} We acknowledge the support from the National Natural Science Foundation of China (Grants Nos. 12274174, T2225013, 52288102 and 12034009).  L.B. thanks the Office of Naval Research (ONR) under Grant No. N00014-17-1-2818 and the Vannevar Bush Faculty Fellowship (VBFF) Grant No. N00014-21-1-2086 from the Department of Defense. L.J.Y. acknowledges the support from the International Center of Future Science, Jilin University. The calculation was performed in the high-performance computing center of Jilin University. The authors thank Prof. P. Liu at Chinese Academy of Sciences (Institute of Metal Research) for the valuable discussion regarding the Hubbard $U$ correction in NaOsO$_3$.\\


\begin{thebibliography}{66}%
\makeatletter
\providecommand \@ifxundefined [1]{%
 \@ifx{#1\undefined}
}%
\providecommand \@ifnum [1]{%
 \ifnum #1\expandafter \@firstoftwo
 \else \expandafter \@secondoftwo
 \fi
}%
\providecommand \@ifx [1]{%
 \ifx #1\expandafter \@firstoftwo
 \else \expandafter \@secondoftwo
 \fi
}%
\providecommand \natexlab [1]{#1}%
\providecommand \enquote  [1]{``#1''}%
\providecommand \bibnamefont  [1]{#1}%
\providecommand \bibfnamefont [1]{#1}%
\providecommand \citenamefont [1]{#1}%
\providecommand \href@noop [0]{\@secondoftwo}%
\providecommand \href [0]{\begingroup \@sanitize@url \@href}%
\providecommand \@href[1]{\@@startlink{#1}\@@href}%
\providecommand \@@href[1]{\endgroup#1\@@endlink}%
\providecommand \@sanitize@url [0]{\catcode `\\12\catcode `\$12\catcode
  `\&12\catcode `\#12\catcode `\^12\catcode `\_12\catcode `\%12\relax}%
\providecommand \@@startlink[1]{}%
\providecommand \@@endlink[0]{}%
\providecommand \url  [0]{\begingroup\@sanitize@url \@url }%
\providecommand \@url [1]{\endgroup\@href {#1}{\urlprefix }}%
\providecommand \urlprefix  [0]{URL }%
\providecommand \Eprint [0]{\href }%
\providecommand \doibase [0]{https://doi.org/}%
\providecommand \selectlanguage [0]{\@gobble}%
\providecommand \bibinfo  [0]{\@secondoftwo}%
\providecommand \bibfield  [0]{\@secondoftwo}%
\providecommand \translation [1]{[#1]}%
\providecommand \BibitemOpen [0]{}%
\providecommand \bibitemStop [0]{}%
\providecommand \bibitemNoStop [0]{.\EOS\space}%
\providecommand \EOS [0]{\spacefactor3000\relax}%
\providecommand \BibitemShut  [1]{\csname bibitem#1\endcsname}%
\let\auto@bib@innerbib\@empty
\bibitem [{\citenamefont {Wolf}\ \emph {et~al.}(2001)\citenamefont {Wolf},
  \citenamefont {Awschalom}, \citenamefont {Buhrman}, \citenamefont {Daughton},
  \citenamefont {von Moln{\'a}r}, \citenamefont {Roukes}, \citenamefont
  {Chtchelkanova},\ and\ \citenamefont {Treger}}]{science2001}%
  \BibitemOpen
  \bibfield  {author} {\bibinfo {author} {\bibfnamefont {S.}~\bibnamefont
  {Wolf}}, \bibinfo {author} {\bibfnamefont {D.}~\bibnamefont {Awschalom}},
  \bibinfo {author} {\bibfnamefont {R.}~\bibnamefont {Buhrman}}, \bibinfo
  {author} {\bibfnamefont {J.}~\bibnamefont {Daughton}}, \bibinfo {author}
  {\bibfnamefont {v.~S.}\ \bibnamefont {von Moln{\'a}r}}, \bibinfo {author}
  {\bibfnamefont {M.}~\bibnamefont {Roukes}}, \bibinfo {author} {\bibfnamefont
  {A.~Y.}\ \bibnamefont {Chtchelkanova}},\ and\ \bibinfo {author}
  {\bibfnamefont {D.}~\bibnamefont {Treger}},\ }\href@noop {} {\bibfield
  {journal} {\bibinfo  {journal} {Science}\ }\textbf {\bibinfo {volume}
  {294}},\ \bibinfo {pages} {1488} (\bibinfo {year} {2001})}\BibitemShut
  {NoStop}%
\bibitem [{\citenamefont {Xia}\ \emph {et~al.}(2011)\citenamefont {Xia},
  \citenamefont {Ge},\ and\ \citenamefont {Chang}}]{spintronic2}%
  \BibitemOpen
  \bibfield  {author} {\bibinfo {author} {\bibfnamefont {J.}~\bibnamefont
  {Xia}}, \bibinfo {author} {\bibfnamefont {W.}~\bibnamefont {Ge}},\ and\
  \bibinfo {author} {\bibfnamefont {K.}~\bibnamefont {Chang}},\ }\href@noop {}
  {\emph {\bibinfo {title} {Semiconductor Spintronics}}}\ (\bibinfo
  {publisher} {{WORLD} {SCIENTIFIC}},\ \bibinfo {year} {2011})\BibitemShut
  {NoStop}%
\bibitem [{\citenamefont {Huang}\ \emph {et~al.}(2021)\citenamefont {Huang},
  \citenamefont {Poloj\"{a}rvi}, \citenamefont {Hiura}, \citenamefont
  {H\"{o}jer}, \citenamefont {Aho}, \citenamefont {Isoaho}, \citenamefont
  {Hakkarainen}, \citenamefont {Guina}, \citenamefont {Sato}, \citenamefont
  {Takayama}, \citenamefont {Murayama}, \citenamefont {Buyanova},\ and\
  \citenamefont {Chen}}]{natphot2021}%
  \BibitemOpen
  \bibfield  {author} {\bibinfo {author} {\bibfnamefont {Y.}~\bibnamefont
  {Huang}}, \bibinfo {author} {\bibfnamefont {V.}~\bibnamefont
  {Poloj\"{a}rvi}}, \bibinfo {author} {\bibfnamefont {S.}~\bibnamefont
  {Hiura}}, \bibinfo {author} {\bibfnamefont {P.}~\bibnamefont {H\"{o}jer}},
  \bibinfo {author} {\bibfnamefont {A.}~\bibnamefont {Aho}}, \bibinfo {author}
  {\bibfnamefont {R.}~\bibnamefont {Isoaho}}, \bibinfo {author} {\bibfnamefont
  {T.}~\bibnamefont {Hakkarainen}}, \bibinfo {author} {\bibfnamefont
  {M.}~\bibnamefont {Guina}}, \bibinfo {author} {\bibfnamefont
  {S.}~\bibnamefont {Sato}}, \bibinfo {author} {\bibfnamefont {J.}~\bibnamefont
  {Takayama}}, \bibinfo {author} {\bibfnamefont {A.}~\bibnamefont {Murayama}},
  \bibinfo {author} {\bibfnamefont {I.~A.}\ \bibnamefont {Buyanova}},\ and\
  \bibinfo {author} {\bibfnamefont {W.~M.}\ \bibnamefont {Chen}},\ }\href@noop
  {} {\bibfield  {journal} {\bibinfo  {journal} {Nat. Photonics}\ }\textbf
  {\bibinfo {volume} {15}},\ \bibinfo {pages} {475} (\bibinfo {year}
  {2021})}\BibitemShut {NoStop}%
\bibitem [{\citenamefont {Li}\ and\ \citenamefont {Yang}(2016)}]{afmnsr}%
  \BibitemOpen
  \bibfield  {author} {\bibinfo {author} {\bibfnamefont {X.}~\bibnamefont
  {Li}}\ and\ \bibinfo {author} {\bibfnamefont {J.}~\bibnamefont {Yang}},\
  }\href@noop {} {\bibfield  {journal} {\bibinfo  {journal} {Natl. Sci. Rev.}\
  }\textbf {\bibinfo {volume} {3}},\ \bibinfo {pages} {365} (\bibinfo {year}
  {2016})}\BibitemShut {NoStop}%
\bibitem [{\citenamefont {N{\v{e}}mec}\ \emph {et~al.}(2018)\citenamefont
  {N{\v{e}}mec}, \citenamefont {Fiebig}, \citenamefont {Kampfrath},\ and\
  \citenamefont {Kimel}}]{afm2018}%
  \BibitemOpen
  \bibfield  {author} {\bibinfo {author} {\bibfnamefont {P.}~\bibnamefont
  {N{\v{e}}mec}}, \bibinfo {author} {\bibfnamefont {M.}~\bibnamefont {Fiebig}},
  \bibinfo {author} {\bibfnamefont {T.}~\bibnamefont {Kampfrath}},\ and\
  \bibinfo {author} {\bibfnamefont {A.~V.}\ \bibnamefont {Kimel}},\ }\href@noop
  {} {\bibfield  {journal} {\bibinfo  {journal} {Nat. Phys.}\ }\textbf
  {\bibinfo {volume} {14}},\ \bibinfo {pages} {229} (\bibinfo {year}
  {2018})}\BibitemShut {NoStop}%
\bibitem [{\citenamefont {Kim}\ \emph {et~al.}(2021)\citenamefont {Kim},
  \citenamefont {Beach}, \citenamefont {Lee}, \citenamefont {Ono},
  \citenamefont {Rasing},\ and\ \citenamefont {Yang}}]{ferri2021}%
  \BibitemOpen
  \bibfield  {author} {\bibinfo {author} {\bibfnamefont {S.~K.}\ \bibnamefont
  {Kim}}, \bibinfo {author} {\bibfnamefont {G.~S.~D.}\ \bibnamefont {Beach}},
  \bibinfo {author} {\bibfnamefont {K.-J.}\ \bibnamefont {Lee}}, \bibinfo
  {author} {\bibfnamefont {T.}~\bibnamefont {Ono}}, \bibinfo {author}
  {\bibfnamefont {T.}~\bibnamefont {Rasing}},\ and\ \bibinfo {author}
  {\bibfnamefont {H.}~\bibnamefont {Yang}},\ }\href@noop {} {\bibfield
  {journal} {\bibinfo  {journal} {Nat. Mater.}\ }\textbf {\bibinfo {volume}
  {21}},\ \bibinfo {pages} {24} (\bibinfo {year} {2021})}\BibitemShut {NoStop}%
\bibitem [{Note1()}]{Note1}%
  \BibitemOpen
  \bibinfo {note} {The asymmetric antiferromagnetic semiconductors and bipolar
  antiferromagnetic semiconductors reported in Refs.~\cite {wang2022,yang2015},
  being ferrimagnetic semiconductors in essence, involve more than one magnetic
  sublattice with almost cancelled magnetic moments. The densities of states of
  these materials (together with ferrimagnetic semiconductors) resemble those
  in ferromagnets, with giant splittings between spin-up and spin-down
  channels. Consequently, the doped carriers in these materials can be highly
  spin-polarized, significantly tuning the magnetic moments.}\BibitemShut
  {Stop}%
\bibitem [{\citenamefont {Wang}\ \emph {et~al.}(2022)\citenamefont {Wang},
  \citenamefont {Wu}, \citenamefont {Zhang},\ and\ \citenamefont
  {Wu}}]{wang2022}%
  \BibitemOpen
  \bibfield  {author} {\bibinfo {author} {\bibfnamefont {P.}~\bibnamefont
  {Wang}}, \bibinfo {author} {\bibfnamefont {D.}~\bibnamefont {Wu}}, \bibinfo
  {author} {\bibfnamefont {K.}~\bibnamefont {Zhang}},\ and\ \bibinfo {author}
  {\bibfnamefont {X.}~\bibnamefont {Wu}},\ }\href@noop {} {\bibfield  {journal}
  {\bibinfo  {journal} {J. Phys. Chem. Lett.}\ }\textbf {\bibinfo {volume}
  {13}},\ \bibinfo {pages} {3850} (\bibinfo {year} {2022})}\BibitemShut
  {NoStop}%
\bibitem [{\citenamefont {Nie}\ and\ \citenamefont {Hu}(2008)}]{ferri2008}%
  \BibitemOpen
  \bibfield  {author} {\bibinfo {author} {\bibfnamefont {Y.-m.}\ \bibnamefont
  {Nie}}\ and\ \bibinfo {author} {\bibfnamefont {X.}~\bibnamefont {Hu}},\
  }\href@noop {} {\bibfield  {journal} {\bibinfo  {journal} {Phys. Rev. Lett.}\
  }\textbf {\bibinfo {volume} {100}},\ \bibinfo {pages} {117203} (\bibinfo
  {year} {2008})}\BibitemShut {NoStop}%
\bibitem [{\citenamefont {Li}\ \emph {et~al.}(2015)\citenamefont {Li},
  \citenamefont {Wu}, \citenamefont {Li},\ and\ \citenamefont
  {Yang}}]{yang2015}%
  \BibitemOpen
  \bibfield  {author} {\bibinfo {author} {\bibfnamefont {X.}~\bibnamefont
  {Li}}, \bibinfo {author} {\bibfnamefont {X.}~\bibnamefont {Wu}}, \bibinfo
  {author} {\bibfnamefont {Z.}~\bibnamefont {Li}},\ and\ \bibinfo {author}
  {\bibfnamefont {J.}~\bibnamefont {Yang}},\ }\href@noop {} {\bibfield
  {journal} {\bibinfo  {journal} {Phys. Rev. B}\ }\textbf {\bibinfo {volume}
  {92}},\ \bibinfo {pages} {125202} (\bibinfo {year} {2015})}\BibitemShut
  {NoStop}%
\bibitem [{\citenamefont {Bousquet}\ and\ \citenamefont
  {Cano}(2016)}]{noncollinear}%
  \BibitemOpen
  \bibfield  {author} {\bibinfo {author} {\bibfnamefont {E.}~\bibnamefont
  {Bousquet}}\ and\ \bibinfo {author} {\bibfnamefont {A.}~\bibnamefont
  {Cano}},\ }\href@noop {} {\bibfield  {journal} {\bibinfo  {journal} {J.
  Phys.: Condens. Matter}\ }\textbf {\bibinfo {volume} {28}},\ \bibinfo {pages}
  {123001} (\bibinfo {year} {2016})}\BibitemShut {NoStop}%
\bibitem [{\citenamefont {Baltz}\ \emph {et~al.}(2018)\citenamefont {Baltz},
  \citenamefont {Manchon}, \citenamefont {Tsoi}, \citenamefont {Moriyama},
  \citenamefont {Ono},\ and\ \citenamefont {Tserkovnyak}}]{afmrmp2018}%
  \BibitemOpen
  \bibfield  {author} {\bibinfo {author} {\bibfnamefont {V.}~\bibnamefont
  {Baltz}}, \bibinfo {author} {\bibfnamefont {A.}~\bibnamefont {Manchon}},
  \bibinfo {author} {\bibfnamefont {M.}~\bibnamefont {Tsoi}}, \bibinfo {author}
  {\bibfnamefont {T.}~\bibnamefont {Moriyama}}, \bibinfo {author}
  {\bibfnamefont {T.}~\bibnamefont {Ono}},\ and\ \bibinfo {author}
  {\bibfnamefont {Y.}~\bibnamefont {Tserkovnyak}},\ }\href@noop {} {\bibfield
  {journal} {\bibinfo  {journal} {Rev. Mod. Phys.}\ }\textbf {\bibinfo {volume}
  {90}},\ \bibinfo {pages} {015005} (\bibinfo {year} {2018})}\BibitemShut
  {NoStop}%
\bibitem [{\citenamefont {Zhou}\ \emph {et~al.}(2020)\citenamefont {Zhou},
  \citenamefont {Marshall}, \citenamefont {Li}, \citenamefont {Li},\ and\
  \citenamefont {He}}]{zhou2020}%
  \BibitemOpen
  \bibfield  {author} {\bibinfo {author} {\bibfnamefont {J.-S.}\ \bibnamefont
  {Zhou}}, \bibinfo {author} {\bibfnamefont {L.~G.}\ \bibnamefont {Marshall}},
  \bibinfo {author} {\bibfnamefont {Z.-Y.}\ \bibnamefont {Li}}, \bibinfo
  {author} {\bibfnamefont {X.}~\bibnamefont {Li}},\ and\ \bibinfo {author}
  {\bibfnamefont {J.-M.}\ \bibnamefont {He}},\ }\href@noop {} {\bibfield
  {journal} {\bibinfo  {journal} {Phys. Rev. B}\ }\textbf {\bibinfo {volume}
  {102}},\ \bibinfo {pages} {104420} (\bibinfo {year} {2020})}\BibitemShut
  {NoStop}%
\bibitem [{\citenamefont {Liu}\ \emph {et~al.}(2023)\citenamefont {Liu},
  \citenamefont {Zhao}, \citenamefont {Bellaiche},\ and\ \citenamefont
  {Ma}}]{zeemanafm}%
  \BibitemOpen
  \bibfield  {author} {\bibinfo {author} {\bibfnamefont {X.}~\bibnamefont
  {Liu}}, \bibinfo {author} {\bibfnamefont {H.~J.}\ \bibnamefont {Zhao}},
  \bibinfo {author} {\bibfnamefont {L.}~\bibnamefont {Bellaiche}},\ and\
  \bibinfo {author} {\bibfnamefont {Y.}~\bibnamefont {Ma}},\ }\href@noop {}
  {\bibfield  {journal} {\bibinfo  {journal} {Submitted}\ } (\bibinfo {year}
  {2023})}\BibitemShut {NoStop}%
\bibitem [{\citenamefont {Liu}\ \emph {et~al.}(2022)\citenamefont {Liu},
  \citenamefont {Wang}, \citenamefont {Zhao}, \citenamefont {Bellaiche},\ and\
  \citenamefont {Ma}}]{liu2022}%
  \BibitemOpen
  \bibfield  {author} {\bibinfo {author} {\bibfnamefont {P.}~\bibnamefont
  {Liu}}, \bibinfo {author} {\bibfnamefont {Y.}~\bibnamefont {Wang}}, \bibinfo
  {author} {\bibfnamefont {H.~J.}\ \bibnamefont {Zhao}}, \bibinfo {author}
  {\bibfnamefont {L.}~\bibnamefont {Bellaiche}},\ and\ \bibinfo {author}
  {\bibfnamefont {Y.}~\bibnamefont {Ma}},\ }\href@noop {} {\bibfield  {journal}
  {\bibinfo  {journal} {Phys. Rev. B}\ }\textbf {\bibinfo {volume} {105}},\
  \bibinfo {pages} {054402} (\bibinfo {year} {2022})}\BibitemShut {NoStop}%
\bibitem [{\citenamefont {Zhou}\ and\ \citenamefont
  {Goodenough}(2005)}]{ytio3}%
  \BibitemOpen
  \bibfield  {author} {\bibinfo {author} {\bibfnamefont {H.~D.}\ \bibnamefont
  {Zhou}}\ and\ \bibinfo {author} {\bibfnamefont {J.~B.}\ \bibnamefont
  {Goodenough}},\ }\href@noop {} {\bibfield  {journal} {\bibinfo  {journal} {J.
  Phys.: Condens. Matter}\ }\textbf {\bibinfo {volume} {17}},\ \bibinfo {pages}
  {7395} (\bibinfo {year} {2005})}\BibitemShut {NoStop}%
\bibitem [{\citenamefont {Disa}\ \emph {et~al.}(2023)\citenamefont {Disa},
  \citenamefont {Curtis}, \citenamefont {Fechner}, \citenamefont {Liu},
  \citenamefont {von Hoegen}, \citenamefont {F\"{o}rst}, \citenamefont {Nova},
  \citenamefont {Narang}, \citenamefont {Maljuk}, \citenamefont {Boris},
  \citenamefont {Keimer},\ and\ \citenamefont {Cavalleri}}]{ytio32}%
  \BibitemOpen
  \bibfield  {author} {\bibinfo {author} {\bibfnamefont {A.~S.}\ \bibnamefont
  {Disa}}, \bibinfo {author} {\bibfnamefont {J.}~\bibnamefont {Curtis}},
  \bibinfo {author} {\bibfnamefont {M.}~\bibnamefont {Fechner}}, \bibinfo
  {author} {\bibfnamefont {A.}~\bibnamefont {Liu}}, \bibinfo {author}
  {\bibfnamefont {A.}~\bibnamefont {von Hoegen}}, \bibinfo {author}
  {\bibfnamefont {M.}~\bibnamefont {F\"{o}rst}}, \bibinfo {author}
  {\bibfnamefont {T.~F.}\ \bibnamefont {Nova}}, \bibinfo {author}
  {\bibfnamefont {P.}~\bibnamefont {Narang}}, \bibinfo {author} {\bibfnamefont
  {A.}~\bibnamefont {Maljuk}}, \bibinfo {author} {\bibfnamefont {A.~V.}\
  \bibnamefont {Boris}}, \bibinfo {author} {\bibfnamefont {B.}~\bibnamefont
  {Keimer}},\ and\ \bibinfo {author} {\bibfnamefont {A.}~\bibnamefont
  {Cavalleri}},\ }\href@noop {} {\bibfield  {journal} {\bibinfo  {journal}
  {Nature}\ }\textbf {\bibinfo {volume} {617}},\ \bibinfo {pages} {73}
  (\bibinfo {year} {2023})}\BibitemShut {NoStop}%
\bibitem [{\citenamefont {Avdeev}\ \emph {et~al.}(2011)\citenamefont {Avdeev},
  \citenamefont {Thorogood}, \citenamefont {Carter}, \citenamefont {Kennedy},
  \citenamefont {Ting}, \citenamefont {Singh},\ and\ \citenamefont
  {Wallwork}}]{catco3}%
  \BibitemOpen
  \bibfield  {author} {\bibinfo {author} {\bibfnamefont {M.}~\bibnamefont
  {Avdeev}}, \bibinfo {author} {\bibfnamefont {G.~J.}\ \bibnamefont
  {Thorogood}}, \bibinfo {author} {\bibfnamefont {M.~L.}\ \bibnamefont
  {Carter}}, \bibinfo {author} {\bibfnamefont {B.~J.}\ \bibnamefont {Kennedy}},
  \bibinfo {author} {\bibfnamefont {J.}~\bibnamefont {Ting}}, \bibinfo {author}
  {\bibfnamefont {D.~J.}\ \bibnamefont {Singh}},\ and\ \bibinfo {author}
  {\bibfnamefont {K.~S.}\ \bibnamefont {Wallwork}},\ }\href@noop {} {\bibfield
  {journal} {\bibinfo  {journal} {J. Am. Chem. Soc.}\ }\textbf {\bibinfo
  {volume} {133}},\ \bibinfo {pages} {1654} (\bibinfo {year}
  {2011})}\BibitemShut {NoStop}%
\bibitem [{\citenamefont {Shi}\ \emph {et~al.}(2009)\citenamefont {Shi},
  \citenamefont {Guo}, \citenamefont {Yu}, \citenamefont {Arai}, \citenamefont
  {Belik}, \citenamefont {Sato}, \citenamefont {Yamaura}, \citenamefont
  {Takayama-Muromachi}, \citenamefont {Tian}, \citenamefont {Yang},
  \citenamefont {Li}, \citenamefont {Varga}, \citenamefont {Mitchell},\ and\
  \citenamefont {Okamoto}}]{naoso3}%
  \BibitemOpen
  \bibfield  {author} {\bibinfo {author} {\bibfnamefont {Y.~G.}\ \bibnamefont
  {Shi}}, \bibinfo {author} {\bibfnamefont {Y.~F.}\ \bibnamefont {Guo}},
  \bibinfo {author} {\bibfnamefont {S.}~\bibnamefont {Yu}}, \bibinfo {author}
  {\bibfnamefont {M.}~\bibnamefont {Arai}}, \bibinfo {author} {\bibfnamefont
  {A.~A.}\ \bibnamefont {Belik}}, \bibinfo {author} {\bibfnamefont
  {A.}~\bibnamefont {Sato}}, \bibinfo {author} {\bibfnamefont {K.}~\bibnamefont
  {Yamaura}}, \bibinfo {author} {\bibfnamefont {E.}~\bibnamefont
  {Takayama-Muromachi}}, \bibinfo {author} {\bibfnamefont {H.~F.}\ \bibnamefont
  {Tian}}, \bibinfo {author} {\bibfnamefont {H.~X.}\ \bibnamefont {Yang}},
  \bibinfo {author} {\bibfnamefont {J.~Q.}\ \bibnamefont {Li}}, \bibinfo
  {author} {\bibfnamefont {T.}~\bibnamefont {Varga}}, \bibinfo {author}
  {\bibfnamefont {J.~F.}\ \bibnamefont {Mitchell}},\ and\ \bibinfo {author}
  {\bibfnamefont {S.}~\bibnamefont {Okamoto}},\ }\href@noop {} {\bibfield
  {journal} {\bibinfo  {journal} {Phys. Rev. B}\ }\textbf {\bibinfo {volume}
  {80}},\ \bibinfo {pages} {161104} (\bibinfo {year} {2009})}\BibitemShut
  {NoStop}%
\bibitem [{\citenamefont {Calder}\ \emph {et~al.}(2012)\citenamefont {Calder},
  \citenamefont {Garlea}, \citenamefont {McMorrow}, \citenamefont {Lumsden},
  \citenamefont {Stone}, \citenamefont {Lang}, \citenamefont {Kim},
  \citenamefont {Schlueter}, \citenamefont {Shi}, \citenamefont {Yamaura},
  \citenamefont {Sun}, \citenamefont {Tsujimoto},\ and\ \citenamefont
  {Christianson}}]{naoso32}%
  \BibitemOpen
  \bibfield  {author} {\bibinfo {author} {\bibfnamefont {S.}~\bibnamefont
  {Calder}}, \bibinfo {author} {\bibfnamefont {V.~O.}\ \bibnamefont {Garlea}},
  \bibinfo {author} {\bibfnamefont {D.~F.}\ \bibnamefont {McMorrow}}, \bibinfo
  {author} {\bibfnamefont {M.~D.}\ \bibnamefont {Lumsden}}, \bibinfo {author}
  {\bibfnamefont {M.~B.}\ \bibnamefont {Stone}}, \bibinfo {author}
  {\bibfnamefont {J.~C.}\ \bibnamefont {Lang}}, \bibinfo {author}
  {\bibfnamefont {J.-W.}\ \bibnamefont {Kim}}, \bibinfo {author} {\bibfnamefont
  {J.~A.}\ \bibnamefont {Schlueter}}, \bibinfo {author} {\bibfnamefont {Y.~G.}\
  \bibnamefont {Shi}}, \bibinfo {author} {\bibfnamefont {K.}~\bibnamefont
  {Yamaura}}, \bibinfo {author} {\bibfnamefont {Y.~S.}\ \bibnamefont {Sun}},
  \bibinfo {author} {\bibfnamefont {Y.}~\bibnamefont {Tsujimoto}},\ and\
  \bibinfo {author} {\bibfnamefont {A.~D.}\ \bibnamefont {Christianson}},\
  }\href@noop {} {\bibfield  {journal} {\bibinfo  {journal} {Phys. Rev. Lett.}\
  }\textbf {\bibinfo {volume} {108}},\ \bibinfo {pages} {257209} (\bibinfo
  {year} {2012})}\BibitemShut {NoStop}%
\bibitem [{\citenamefont {Sereika}\ \emph {et~al.}(2020)\citenamefont
  {Sereika}, \citenamefont {Liu}, \citenamefont {Kim}, \citenamefont {Kim},
  \citenamefont {Zhang}, \citenamefont {Chen}, \citenamefont {Yamaura},
  \citenamefont {Park}, \citenamefont {Franchini}, \citenamefont {Ding},\ and\
  \citenamefont {k.~Mao}}]{naoso33}%
  \BibitemOpen
  \bibfield  {author} {\bibinfo {author} {\bibfnamefont {R.}~\bibnamefont
  {Sereika}}, \bibinfo {author} {\bibfnamefont {P.}~\bibnamefont {Liu}},
  \bibinfo {author} {\bibfnamefont {B.}~\bibnamefont {Kim}}, \bibinfo {author}
  {\bibfnamefont {S.}~\bibnamefont {Kim}}, \bibinfo {author} {\bibfnamefont
  {J.}~\bibnamefont {Zhang}}, \bibinfo {author} {\bibfnamefont
  {B.}~\bibnamefont {Chen}}, \bibinfo {author} {\bibfnamefont {K.}~\bibnamefont
  {Yamaura}}, \bibinfo {author} {\bibfnamefont {C.}~\bibnamefont {Park}},
  \bibinfo {author} {\bibfnamefont {C.}~\bibnamefont {Franchini}}, \bibinfo
  {author} {\bibfnamefont {Y.}~\bibnamefont {Ding}},\ and\ \bibinfo {author}
  {\bibfnamefont {H.}~\bibnamefont {k.~Mao}},\ }\href@noop {} {\bibfield
  {journal} {\bibinfo  {journal} {npj Quantum Mater.}\ }\textbf {\bibinfo
  {volume} {5}},\ \bibinfo {pages} {66} (\bibinfo {year} {2020})}\BibitemShut
  {NoStop}%
\bibitem [{\citenamefont {Vale}\ \emph {et~al.}(2018)\citenamefont {Vale},
  \citenamefont {Calder}, \citenamefont {Donnerer}, \citenamefont {Pincini},
  \citenamefont {Shi}, \citenamefont {Tsujimoto}, \citenamefont {Yamaura},
  \citenamefont {Sala}, \citenamefont {van~den Brink}, \citenamefont
  {Christianson},\ and\ \citenamefont {McMorrow}}]{naoso3mit}%
  \BibitemOpen
  \bibfield  {author} {\bibinfo {author} {\bibfnamefont {J.~G.}\ \bibnamefont
  {Vale}}, \bibinfo {author} {\bibfnamefont {S.}~\bibnamefont {Calder}},
  \bibinfo {author} {\bibfnamefont {C.}~\bibnamefont {Donnerer}}, \bibinfo
  {author} {\bibfnamefont {D.}~\bibnamefont {Pincini}}, \bibinfo {author}
  {\bibfnamefont {Y.~G.}\ \bibnamefont {Shi}}, \bibinfo {author} {\bibfnamefont
  {Y.}~\bibnamefont {Tsujimoto}}, \bibinfo {author} {\bibfnamefont
  {K.}~\bibnamefont {Yamaura}}, \bibinfo {author} {\bibfnamefont {M.~M.}\
  \bibnamefont {Sala}}, \bibinfo {author} {\bibfnamefont {J.}~\bibnamefont
  {van~den Brink}}, \bibinfo {author} {\bibfnamefont {A.~D.}\ \bibnamefont
  {Christianson}},\ and\ \bibinfo {author} {\bibfnamefont {D.~F.}\ \bibnamefont
  {McMorrow}},\ }\href@noop {} {\bibfield  {journal} {\bibinfo  {journal}
  {Phys. Rev. Lett.}\ }\textbf {\bibinfo {volume} {120}},\ \bibinfo {pages}
  {227203} (\bibinfo {year} {2018})}\BibitemShut {NoStop}%
\bibitem [{\citenamefont {Franchini}\ \emph {et~al.}(2011)\citenamefont
  {Franchini}, \citenamefont {Archer}, \citenamefont {He}, \citenamefont
  {Chen}, \citenamefont {Filippetti},\ and\ \citenamefont
  {Sanvito}}]{catco3prb}%
  \BibitemOpen
  \bibfield  {author} {\bibinfo {author} {\bibfnamefont {C.}~\bibnamefont
  {Franchini}}, \bibinfo {author} {\bibfnamefont {T.}~\bibnamefont {Archer}},
  \bibinfo {author} {\bibfnamefont {J.}~\bibnamefont {He}}, \bibinfo {author}
  {\bibfnamefont {X.-Q.}\ \bibnamefont {Chen}}, \bibinfo {author}
  {\bibfnamefont {A.}~\bibnamefont {Filippetti}},\ and\ \bibinfo {author}
  {\bibfnamefont {S.}~\bibnamefont {Sanvito}},\ }\href@noop {} {\bibfield
  {journal} {\bibinfo  {journal} {Phys. Rev. B}\ }\textbf {\bibinfo {volume}
  {83}},\ \bibinfo {pages} {220402} (\bibinfo {year} {2011})}\BibitemShut
  {NoStop}%
\bibitem [{\citenamefont {Wang}\ \emph {et~al.}(2015)\citenamefont {Wang},
  \citenamefont {He},\ and\ \citenamefont {Wu}}]{ctou2}%
  \BibitemOpen
  \bibfield  {author} {\bibinfo {author} {\bibfnamefont {H.}~\bibnamefont
  {Wang}}, \bibinfo {author} {\bibfnamefont {L.}~\bibnamefont {He}},\ and\
  \bibinfo {author} {\bibfnamefont {X.}~\bibnamefont {Wu}},\ }\href@noop {}
  {\bibfield  {journal} {\bibinfo  {journal} {Comput. Mater. Sci.}\ }\textbf
  {\bibinfo {volume} {96}},\ \bibinfo {pages} {171} (\bibinfo {year}
  {2015})}\BibitemShut {NoStop}%
\bibitem [{\citenamefont {Tao}\ and\ \citenamefont
  {Tsymbal}(2021)}]{spinsplitting}%
  \BibitemOpen
  \bibfield  {author} {\bibinfo {author} {\bibfnamefont {L.~L.}\ \bibnamefont
  {Tao}}\ and\ \bibinfo {author} {\bibfnamefont {E.~Y.}\ \bibnamefont
  {Tsymbal}},\ }\href@noop {} {\bibfield  {journal} {\bibinfo  {journal} {J.
  Phys. D: Appl. Phys.}\ }\textbf {\bibinfo {volume} {54}},\ \bibinfo {pages}
  {113001} (\bibinfo {year} {2021})}\BibitemShut {NoStop}%
\bibitem [{\citenamefont {Manchon}\ \emph {et~al.}(2015)\citenamefont
  {Manchon}, \citenamefont {Koo}, \citenamefont {Nitta}, \citenamefont
  {Frolov},\ and\ \citenamefont {Duine}}]{spinsplitting2}%
  \BibitemOpen
  \bibfield  {author} {\bibinfo {author} {\bibfnamefont {A.}~\bibnamefont
  {Manchon}}, \bibinfo {author} {\bibfnamefont {H.~C.}\ \bibnamefont {Koo}},
  \bibinfo {author} {\bibfnamefont {J.}~\bibnamefont {Nitta}}, \bibinfo
  {author} {\bibfnamefont {S.~M.}\ \bibnamefont {Frolov}},\ and\ \bibinfo
  {author} {\bibfnamefont {R.~A.}\ \bibnamefont {Duine}},\ }\href@noop {}
  {\bibfield  {journal} {\bibinfo  {journal} {Nat. Mater.}\ }\textbf {\bibinfo
  {volume} {14}},\ \bibinfo {pages} {871} (\bibinfo {year} {2015})}\BibitemShut
  {NoStop}%
\bibitem [{\citenamefont {Bychkov}\ and\ \citenamefont
  {Rashba}(1984)}]{rashbajetp}%
  \BibitemOpen
  \bibfield  {author} {\bibinfo {author} {\bibfnamefont {Y.~A.}\ \bibnamefont
  {Bychkov}}\ and\ \bibinfo {author} {\bibfnamefont {E.~I.}\ \bibnamefont
  {Rashba}},\ }\href@noop {} {\bibfield  {journal} {\bibinfo  {journal} {JETP
  Lett.}\ }\textbf {\bibinfo {volume} {39}},\ \bibinfo {pages} {78} (\bibinfo
  {year} {1984})}\BibitemShut {NoStop}%
\bibitem [{\citenamefont {Dresselhaus}(1955)}]{dresselhausoriginal}%
  \BibitemOpen
  \bibfield  {author} {\bibinfo {author} {\bibfnamefont {G.}~\bibnamefont
  {Dresselhaus}},\ }\href@noop {} {\bibfield  {journal} {\bibinfo  {journal}
  {Phys. Rev.}\ }\textbf {\bibinfo {volume} {100}},\ \bibinfo {pages} {580}
  (\bibinfo {year} {1955})}\BibitemShut {NoStop}%
\bibitem [{\citenamefont {Sch\"{a}pers}(2016)}]{spintronic}%
  \BibitemOpen
  \bibfield  {author} {\bibinfo {author} {\bibfnamefont {T.}~\bibnamefont
  {Sch\"{a}pers}},\ }\href@noop {} {\emph {\bibinfo {title} {Semiconductor
  Spintronics}}}\ (\bibinfo  {publisher} {De Gruyter},\ \bibinfo {year}
  {2016})\BibitemShut {NoStop}%
\bibitem [{\citenamefont {Yuan}\ \emph
  {et~al.}(2021{\natexlab{a}})\citenamefont {Yuan}, \citenamefont {Wang},
  \citenamefont {Luo},\ and\ \citenamefont {Zunger}}]{zungerspinsplitting}%
  \BibitemOpen
  \bibfield  {author} {\bibinfo {author} {\bibfnamefont {L.-D.}\ \bibnamefont
  {Yuan}}, \bibinfo {author} {\bibfnamefont {Z.}~\bibnamefont {Wang}}, \bibinfo
  {author} {\bibfnamefont {J.-W.}\ \bibnamefont {Luo}},\ and\ \bibinfo {author}
  {\bibfnamefont {A.}~\bibnamefont {Zunger}},\ }\href@noop {} {\bibfield
  {journal} {\bibinfo  {journal} {Phys. Rev. Mater.}\ }\textbf {\bibinfo
  {volume} {5}},\ \bibinfo {pages} {014409} (\bibinfo {year}
  {2021}{\natexlab{a}})}\BibitemShut {NoStop}%
\bibitem [{\citenamefont {Yuan}\ \emph
  {et~al.}(2021{\natexlab{b}})\citenamefont {Yuan}, \citenamefont {Wang},
  \citenamefont {Luo},\ and\ \citenamefont {Zunger}}]{zungerspinsplitting2}%
  \BibitemOpen
  \bibfield  {author} {\bibinfo {author} {\bibfnamefont {L.-D.}\ \bibnamefont
  {Yuan}}, \bibinfo {author} {\bibfnamefont {Z.}~\bibnamefont {Wang}}, \bibinfo
  {author} {\bibfnamefont {J.-W.}\ \bibnamefont {Luo}},\ and\ \bibinfo {author}
  {\bibfnamefont {A.}~\bibnamefont {Zunger}},\ }\href@noop {} {\bibfield
  {journal} {\bibinfo  {journal} {Phys. Rev. B}\ }\textbf {\bibinfo {volume}
  {103}},\ \bibinfo {pages} {224410} (\bibinfo {year}
  {2021}{\natexlab{b}})}\BibitemShut {NoStop}%
\bibitem [{\citenamefont {Dang}\ \emph {et~al.}(2022)\citenamefont {Dang},
  \citenamefont {Zhu}, \citenamefont {Zhu}, \citenamefont {Chen}, \citenamefont
  {Song},\ and\ \citenamefont {Qi}}]{qi2022}%
  \BibitemOpen
  \bibfield  {author} {\bibinfo {author} {\bibfnamefont {W.}~\bibnamefont
  {Dang}}, \bibinfo {author} {\bibfnamefont {M.}~\bibnamefont {Zhu}}, \bibinfo
  {author} {\bibfnamefont {Z.}~\bibnamefont {Zhu}}, \bibinfo {author}
  {\bibfnamefont {X.}~\bibnamefont {Chen}}, \bibinfo {author} {\bibfnamefont
  {Z.}~\bibnamefont {Song}},\ and\ \bibinfo {author} {\bibfnamefont
  {J.}~\bibnamefont {Qi}},\ }\href@noop {} {\bibfield  {journal} {\bibinfo
  {journal} {Phys. Rev. Appl.}\ }\textbf {\bibinfo {volume} {18}},\ \bibinfo
  {pages} {064086} (\bibinfo {year} {2022})}\BibitemShut {NoStop}%
\bibitem [{\citenamefont {Zhao}\ \emph {et~al.}(2022)\citenamefont {Zhao},
  \citenamefont {Liu}, \citenamefont {Wang}, \citenamefont {Yang},
  \citenamefont {Bellaiche},\ and\ \citenamefont {Ma}}]{zhao2022zeeman}%
  \BibitemOpen
  \bibfield  {author} {\bibinfo {author} {\bibfnamefont {H.~J.}\ \bibnamefont
  {Zhao}}, \bibinfo {author} {\bibfnamefont {X.}~\bibnamefont {Liu}}, \bibinfo
  {author} {\bibfnamefont {Y.}~\bibnamefont {Wang}}, \bibinfo {author}
  {\bibfnamefont {Y.}~\bibnamefont {Yang}}, \bibinfo {author} {\bibfnamefont
  {L.}~\bibnamefont {Bellaiche}},\ and\ \bibinfo {author} {\bibfnamefont
  {Y.}~\bibnamefont {Ma}},\ }\href@noop {} {\bibfield  {journal} {\bibinfo
  {journal} {Phys. Rev. Lett.}\ }\textbf {\bibinfo {volume} {129}},\ \bibinfo
  {pages} {187602} (\bibinfo {year} {2022})}\BibitemShut {NoStop}%
\bibitem [{Note2()}]{Note2}%
  \BibitemOpen
  \bibinfo {note} {In semiconductors, electron doping or hole doping occurs
  around the CBM and VBM, respectively.}\BibitemShut {Stop}%
\bibitem [{\citenamefont {White}(1969)}]{rfeo3}%
  \BibitemOpen
  \bibfield  {author} {\bibinfo {author} {\bibfnamefont {R.~L.}\ \bibnamefont
  {White}},\ }\href@noop {} {\bibfield  {journal} {\bibinfo  {journal} {J.
  Appl. Phys.}\ }\textbf {\bibinfo {volume} {40}},\ \bibinfo {pages} {1061}
  (\bibinfo {year} {1969})}\BibitemShut {NoStop}%
\bibitem [{\citenamefont {Rajeswaran}\ \emph {et~al.}(2012)\citenamefont
  {Rajeswaran}, \citenamefont {Khomskii}, \citenamefont {Zvezdin},
  \citenamefont {Rao},\ and\ \citenamefont {Sundaresan}}]{rcro3}%
  \BibitemOpen
  \bibfield  {author} {\bibinfo {author} {\bibfnamefont {B.}~\bibnamefont
  {Rajeswaran}}, \bibinfo {author} {\bibfnamefont {D.~I.}\ \bibnamefont
  {Khomskii}}, \bibinfo {author} {\bibfnamefont {A.~K.}\ \bibnamefont
  {Zvezdin}}, \bibinfo {author} {\bibfnamefont {C.~N.~R.}\ \bibnamefont
  {Rao}},\ and\ \bibinfo {author} {\bibfnamefont {A.}~\bibnamefont
  {Sundaresan}},\ }\href@noop {} {\bibfield  {journal} {\bibinfo  {journal}
  {Phys. Rev. B}\ }\textbf {\bibinfo {volume} {86}},\ \bibinfo {pages} {214409}
  (\bibinfo {year} {2012})}\BibitemShut {NoStop}%
\bibitem [{\citenamefont {Rodriguez}\ \emph {et~al.}(2011)\citenamefont
  {Rodriguez}, \citenamefont {Poineau}, \citenamefont {Llobet}, \citenamefont
  {Kennedy}, \citenamefont {Avdeev}, \citenamefont {Thorogood}, \citenamefont
  {Carter}, \citenamefont {Seshadri}, \citenamefont {Singh},\ and\
  \citenamefont {Cheetham}}]{srtco3}%
  \BibitemOpen
  \bibfield  {author} {\bibinfo {author} {\bibfnamefont {E.~E.}\ \bibnamefont
  {Rodriguez}}, \bibinfo {author} {\bibfnamefont {F.}~\bibnamefont {Poineau}},
  \bibinfo {author} {\bibfnamefont {A.}~\bibnamefont {Llobet}}, \bibinfo
  {author} {\bibfnamefont {B.~J.}\ \bibnamefont {Kennedy}}, \bibinfo {author}
  {\bibfnamefont {M.}~\bibnamefont {Avdeev}}, \bibinfo {author} {\bibfnamefont
  {G.~J.}\ \bibnamefont {Thorogood}}, \bibinfo {author} {\bibfnamefont {M.~L.}\
  \bibnamefont {Carter}}, \bibinfo {author} {\bibfnamefont {R.}~\bibnamefont
  {Seshadri}}, \bibinfo {author} {\bibfnamefont {D.~J.}\ \bibnamefont
  {Singh}},\ and\ \bibinfo {author} {\bibfnamefont {A.~K.}\ \bibnamefont
  {Cheetham}},\ }\href@noop {} {\bibfield  {journal} {\bibinfo  {journal}
  {Phys. Rev. Lett.}\ }\textbf {\bibinfo {volume} {106}},\ \bibinfo {pages}
  {067201} (\bibinfo {year} {2011})}\BibitemShut {NoStop}%
\bibitem [{\citenamefont {Bellaiche}\ \emph {et~al.}(2012)\citenamefont
  {Bellaiche}, \citenamefont {Gui},\ and\ \citenamefont
  {Kornev}}]{bellaiche2012}%
  \BibitemOpen
  \bibfield  {author} {\bibinfo {author} {\bibfnamefont {L.}~\bibnamefont
  {Bellaiche}}, \bibinfo {author} {\bibfnamefont {Z.}~\bibnamefont {Gui}},\
  and\ \bibinfo {author} {\bibfnamefont {I.~A.}\ \bibnamefont {Kornev}},\
  }\href@noop {} {\bibfield  {journal} {\bibinfo  {journal} {J. Phys.: Condens.
  Matter}\ }\textbf {\bibinfo {volume} {24}},\ \bibinfo {pages} {312201}
  (\bibinfo {year} {2012})}\BibitemShut {NoStop}%
\bibitem [{\citenamefont {Zhao}\ \emph {et~al.}(2021)\citenamefont {Zhao},
  \citenamefont {Chen}, \citenamefont {Prosandeev}, \citenamefont {Paillard},
  \citenamefont {Patel}, \citenamefont {{\'{I}}{\~{n}}iguez},\ and\
  \citenamefont {Bellaiche}}]{zhao2021}%
  \BibitemOpen
  \bibfield  {author} {\bibinfo {author} {\bibfnamefont {H.~J.}\ \bibnamefont
  {Zhao}}, \bibinfo {author} {\bibfnamefont {P.}~\bibnamefont {Chen}}, \bibinfo
  {author} {\bibfnamefont {S.}~\bibnamefont {Prosandeev}}, \bibinfo {author}
  {\bibfnamefont {C.}~\bibnamefont {Paillard}}, \bibinfo {author}
  {\bibfnamefont {K.}~\bibnamefont {Patel}}, \bibinfo {author} {\bibfnamefont
  {J.}~\bibnamefont {{\'{I}}{\~{n}}iguez}},\ and\ \bibinfo {author}
  {\bibfnamefont {L.}~\bibnamefont {Bellaiche}},\ }\href@noop {} {\bibfield
  {journal} {\bibinfo  {journal} {Adv. Elect. Mater.}\ }\textbf {\bibinfo
  {volume} {8}},\ \bibinfo {pages} {2100639} (\bibinfo {year}
  {2021})}\BibitemShut {NoStop}%
\bibitem [{Note3()}]{Note3}%
  \BibitemOpen
  \bibinfo {note} {See Supplementary Material at {\protect \color {blue}a link}
  for the methods used in the present work, and some numerical results
  regarding NaOsO$_3$, CaTcO$_3$ and YFeO$_3$.}\BibitemShut {Stop}%
\bibitem [{\citenamefont {Kresse}\ and\ \citenamefont
  {Furthm{\"u}ller}(1996)}]{kresse1996efficient}%
  \BibitemOpen
  \bibfield  {author} {\bibinfo {author} {\bibfnamefont {G.}~\bibnamefont
  {Kresse}}\ and\ \bibinfo {author} {\bibfnamefont {J.}~\bibnamefont
  {Furthm{\"u}ller}},\ }\href@noop {} {\bibfield  {journal} {\bibinfo
  {journal} {Phys. Rev. B}\ }\textbf {\bibinfo {volume} {54}},\ \bibinfo
  {pages} {11169} (\bibinfo {year} {1996})}\BibitemShut {NoStop}%
\bibitem [{\citenamefont {Kresse}\ and\ \citenamefont
  {Joubert}(1999)}]{kresse1999ultrasoft}%
  \BibitemOpen
  \bibfield  {author} {\bibinfo {author} {\bibfnamefont {G.}~\bibnamefont
  {Kresse}}\ and\ \bibinfo {author} {\bibfnamefont {D.}~\bibnamefont
  {Joubert}},\ }\href@noop {} {\bibfield  {journal} {\bibinfo  {journal} {Phys.
  Rev. B}\ }\textbf {\bibinfo {volume} {59}},\ \bibinfo {pages} {1758}
  (\bibinfo {year} {1999})}\BibitemShut {NoStop}%
\bibitem [{\citenamefont {Bl{\"o}chl}(1994)}]{blochl1994projector}%
  \BibitemOpen
  \bibfield  {author} {\bibinfo {author} {\bibfnamefont {P.~E.}\ \bibnamefont
  {Bl{\"o}chl}},\ }\href@noop {} {\bibfield  {journal} {\bibinfo  {journal}
  {Phys. Rev. B}\ }\textbf {\bibinfo {volume} {50}},\ \bibinfo {pages} {17953}
  (\bibinfo {year} {1994})}\BibitemShut {NoStop}%
\bibitem [{\citenamefont {Ceperley}\ and\ \citenamefont {Alder}(1980)}]{ldaca}%
  \BibitemOpen
  \bibfield  {author} {\bibinfo {author} {\bibfnamefont {D.~M.}\ \bibnamefont
  {Ceperley}}\ and\ \bibinfo {author} {\bibfnamefont {B.~J.}\ \bibnamefont
  {Alder}},\ }\href@noop {} {\bibfield  {journal} {\bibinfo  {journal} {Phys.
  Rev. Lett.}\ }\textbf {\bibinfo {volume} {45}},\ \bibinfo {pages} {566}
  (\bibinfo {year} {1980})}\BibitemShut {NoStop}%
\bibitem [{\citenamefont {Dudarev}\ \emph {et~al.}(1998)\citenamefont
  {Dudarev}, \citenamefont {Botton}, \citenamefont {Savrasov}, \citenamefont
  {Humphreys},\ and\ \citenamefont {Sutton}}]{ldau}%
  \BibitemOpen
  \bibfield  {author} {\bibinfo {author} {\bibfnamefont {S.~L.}\ \bibnamefont
  {Dudarev}}, \bibinfo {author} {\bibfnamefont {G.~A.}\ \bibnamefont {Botton}},
  \bibinfo {author} {\bibfnamefont {S.~Y.}\ \bibnamefont {Savrasov}}, \bibinfo
  {author} {\bibfnamefont {C.~J.}\ \bibnamefont {Humphreys}},\ and\ \bibinfo
  {author} {\bibfnamefont {A.~P.}\ \bibnamefont {Sutton}},\ }\href@noop {}
  {\bibfield  {journal} {\bibinfo  {journal} {Phys. Rev. B}\ }\textbf {\bibinfo
  {volume} {57}},\ \bibinfo {pages} {1505} (\bibinfo {year}
  {1998})}\BibitemShut {NoStop}%
\bibitem [{\citenamefont {Zhang}\ and\ \citenamefont {Tong}(2012)}]{ctojpcm}%
  \BibitemOpen
  \bibfield  {author} {\bibinfo {author} {\bibfnamefont {W.}~\bibnamefont
  {Zhang}}\ and\ \bibinfo {author} {\bibfnamefont {P.}~\bibnamefont {Tong}},\
  }\href@noop {} {\bibfield  {journal} {\bibinfo  {journal} {J. Phys.: Condens.
  Matter}\ }\textbf {\bibinfo {volume} {24}},\ \bibinfo {pages} {185401}
  (\bibinfo {year} {2012})}\BibitemShut {NoStop}%
\bibitem [{\citenamefont {Zhao}\ \emph {et~al.}(2017)\citenamefont {Zhao},
  \citenamefont {Bellaiche}, \citenamefont {Chen},\ and\ \citenamefont
  {{\'{I}}{\~{n}}iguez}}]{gfou3}%
  \BibitemOpen
  \bibfield  {author} {\bibinfo {author} {\bibfnamefont {H.~J.}\ \bibnamefont
  {Zhao}}, \bibinfo {author} {\bibfnamefont {L.}~\bibnamefont {Bellaiche}},
  \bibinfo {author} {\bibfnamefont {X.~M.}\ \bibnamefont {Chen}},\ and\
  \bibinfo {author} {\bibfnamefont {J.}~\bibnamefont {{\'{I}}{\~{n}}iguez}},\
  }\href@noop {} {\bibfield  {journal} {\bibinfo  {journal} {Nat. Commun.}\
  }\textbf {\bibinfo {volume} {8}},\ \bibinfo {pages} {14025} (\bibinfo {year}
  {2017})}\BibitemShut {NoStop}%
\bibitem [{\citenamefont {Momma}\ and\ \citenamefont
  {Izumi}(2011)}]{momma2011vesta}%
  \BibitemOpen
  \bibfield  {author} {\bibinfo {author} {\bibfnamefont {K.}~\bibnamefont
  {Momma}}\ and\ \bibinfo {author} {\bibfnamefont {F.}~\bibnamefont {Izumi}},\
  }\href@noop {} {\bibfield  {journal} {\bibinfo  {journal} {J. Appl.
  Crystallogr.}\ }\textbf {\bibinfo {volume} {44}},\ \bibinfo {pages} {1272}
  (\bibinfo {year} {2011})}\BibitemShut {NoStop}%
\bibitem [{see()}]{seekpath}%
  \BibitemOpen
  \href@noop {} {\emph {\bibinfo {title} {SeeK-path}}}\ (\bibinfo  {publisher}
  {https://www.materialscloud.org/work/tools/seekpath})\BibitemShut {NoStop}%
\bibitem [{\citenamefont {Hinuma}\ \emph {et~al.}(2017)\citenamefont {Hinuma},
  \citenamefont {Pizzi}, \citenamefont {Kumagai}, \citenamefont {Oba},\ and\
  \citenamefont {Tanaka}}]{seekpath2}%
  \BibitemOpen
  \bibfield  {author} {\bibinfo {author} {\bibfnamefont {Y.}~\bibnamefont
  {Hinuma}}, \bibinfo {author} {\bibfnamefont {G.}~\bibnamefont {Pizzi}},
  \bibinfo {author} {\bibfnamefont {Y.}~\bibnamefont {Kumagai}}, \bibinfo
  {author} {\bibfnamefont {F.}~\bibnamefont {Oba}},\ and\ \bibinfo {author}
  {\bibfnamefont {I.}~\bibnamefont {Tanaka}},\ }\href
  {https://doi.org/10.1016/j.commatsci.2016.10.015} {\bibfield  {journal}
  {\bibinfo  {journal} {Comp. Mater. Sci.}\ }\textbf {\bibinfo {volume}
  {128}},\ \bibinfo {pages} {140} (\bibinfo {year} {2017})}\BibitemShut
  {NoStop}%
\bibitem [{\citenamefont {Togo}\ and\ \citenamefont
  {Tanaka}(2018)}]{seekpath3}%
  \BibitemOpen
  \bibfield  {author} {\bibinfo {author} {\bibfnamefont {A.}~\bibnamefont
  {Togo}}\ and\ \bibinfo {author} {\bibfnamefont {I.}~\bibnamefont {Tanaka}},\
  }\href@noop {} {\bibinfo {title} {$\texttt{Spglib}$: a software library for
  crystal symmetry search}} (\bibinfo {year} {2018}),\ \Eprint
  {https://arxiv.org/abs/arXiv:1808.01590} {arXiv:1808.01590} \BibitemShut
  {NoStop}%
\bibitem [{\citenamefont {Herath}\ \emph {et~al.}(2020)\citenamefont {Herath},
  \citenamefont {Tavadze}, \citenamefont {He}, \citenamefont {Bousquet},
  \citenamefont {Singh}, \citenamefont {Munoz},\ and\ \citenamefont
  {Romero}}]{pyprocar}%
  \BibitemOpen
  \bibfield  {author} {\bibinfo {author} {\bibfnamefont {U.}~\bibnamefont
  {Herath}}, \bibinfo {author} {\bibfnamefont {P.}~\bibnamefont {Tavadze}},
  \bibinfo {author} {\bibfnamefont {X.}~\bibnamefont {He}}, \bibinfo {author}
  {\bibfnamefont {E.}~\bibnamefont {Bousquet}}, \bibinfo {author}
  {\bibfnamefont {S.}~\bibnamefont {Singh}}, \bibinfo {author} {\bibfnamefont
  {F.}~\bibnamefont {Munoz}},\ and\ \bibinfo {author} {\bibfnamefont {A.~H.}\
  \bibnamefont {Romero}},\ }\href@noop {} {\bibfield  {journal} {\bibinfo
  {journal} {Comput. Phys. Commun.}\ }\textbf {\bibinfo {volume} {251}},\
  \bibinfo {pages} {107080} (\bibinfo {year} {2020})}\BibitemShut {NoStop}%
\bibitem [{pyp()}]{pyprocar2}%
  \BibitemOpen
  \href@noop {} {\emph {\bibinfo {title} {PYPROCAR}}}\ (\bibinfo  {publisher}
  {https://romerogroup.github.io/pyprocar/index.html})\BibitemShut {NoStop}%
\bibitem [{\citenamefont {Hunter}(2007)}]{hunter2007matplotlib}%
  \BibitemOpen
  \bibfield  {author} {\bibinfo {author} {\bibfnamefont {J.~D.}\ \bibnamefont
  {Hunter}},\ }\href@noop {} {\bibfield  {journal} {\bibinfo  {journal}
  {Comput. Sci. Eng.}\ }\textbf {\bibinfo {volume} {9}},\ \bibinfo {pages} {90}
  (\bibinfo {year} {2007})}\BibitemShut {NoStop}%
\bibitem [{mag()}]{magndata}%
  \BibitemOpen
  \href@noop {} {\emph {\bibinfo {title} {MAGNDATA}}}\ (\bibinfo  {publisher}
  {http://webbdcrista1.ehu.es/magndata})\BibitemShut {NoStop}%
\bibitem [{\citenamefont {Gallego}\ \emph
  {et~al.}(2016{\natexlab{a}})\citenamefont {Gallego}, \citenamefont
  {Perez-Mato}, \citenamefont {Elcoro}, \citenamefont {Tasci}, \citenamefont
  {Hanson}, \citenamefont {Momma}, \citenamefont {Aroyo},\ and\ \citenamefont
  {Madariaga}}]{magndata2}%
  \BibitemOpen
  \bibfield  {author} {\bibinfo {author} {\bibfnamefont {S.~V.}\ \bibnamefont
  {Gallego}}, \bibinfo {author} {\bibfnamefont {J.~M.}\ \bibnamefont
  {Perez-Mato}}, \bibinfo {author} {\bibfnamefont {L.}~\bibnamefont {Elcoro}},
  \bibinfo {author} {\bibfnamefont {E.~S.}\ \bibnamefont {Tasci}}, \bibinfo
  {author} {\bibfnamefont {R.~M.}\ \bibnamefont {Hanson}}, \bibinfo {author}
  {\bibfnamefont {K.}~\bibnamefont {Momma}}, \bibinfo {author} {\bibfnamefont
  {M.~I.}\ \bibnamefont {Aroyo}},\ and\ \bibinfo {author} {\bibfnamefont
  {G.}~\bibnamefont {Madariaga}},\ }\href@noop {} {\bibfield  {journal}
  {\bibinfo  {journal} {J. Appl. Crystallogr.}\ }\textbf {\bibinfo {volume}
  {49}},\ \bibinfo {pages} {1750} (\bibinfo {year}
  {2016}{\natexlab{a}})}\BibitemShut {NoStop}%
\bibitem [{\citenamefont {Gallego}\ \emph
  {et~al.}(2016{\natexlab{b}})\citenamefont {Gallego}, \citenamefont
  {Perez-Mato}, \citenamefont {Elcoro}, \citenamefont {Tasci}, \citenamefont
  {Hanson}, \citenamefont {Aroyo},\ and\ \citenamefont
  {Madariaga}}]{magndata3}%
  \BibitemOpen
  \bibfield  {author} {\bibinfo {author} {\bibfnamefont {S.~V.}\ \bibnamefont
  {Gallego}}, \bibinfo {author} {\bibfnamefont {J.~M.}\ \bibnamefont
  {Perez-Mato}}, \bibinfo {author} {\bibfnamefont {L.}~\bibnamefont {Elcoro}},
  \bibinfo {author} {\bibfnamefont {E.~S.}\ \bibnamefont {Tasci}}, \bibinfo
  {author} {\bibfnamefont {R.~M.}\ \bibnamefont {Hanson}}, \bibinfo {author}
  {\bibfnamefont {M.~I.}\ \bibnamefont {Aroyo}},\ and\ \bibinfo {author}
  {\bibfnamefont {G.}~\bibnamefont {Madariaga}},\ }\href@noop {} {\bibfield
  {journal} {\bibinfo  {journal} {J. Appl. Crystallogr.}\ }\textbf {\bibinfo
  {volume} {49}},\ \bibinfo {pages} {1941} (\bibinfo {year}
  {2016}{\natexlab{b}})}\BibitemShut {NoStop}%
\bibitem [{\citenamefont {Ryee}\ and\ \citenamefont
  {Han}(2018)}]{jpcmnoncollinear}%
  \BibitemOpen
  \bibfield  {author} {\bibinfo {author} {\bibfnamefont {S.}~\bibnamefont
  {Ryee}}\ and\ \bibinfo {author} {\bibfnamefont {M.~J.}\ \bibnamefont {Han}},\
  }\href@noop {} {\bibfield  {journal} {\bibinfo  {journal} {J. Phys.: Condens.
  Matter}\ }\textbf {\bibinfo {volume} {30}},\ \bibinfo {pages} {275802}
  (\bibinfo {year} {2018})}\BibitemShut {NoStop}%
\bibitem [{\citenamefont {Bousquet}\ and\ \citenamefont
  {Spaldin}(2010)}]{exchangej}%
  \BibitemOpen
  \bibfield  {author} {\bibinfo {author} {\bibfnamefont {E.}~\bibnamefont
  {Bousquet}}\ and\ \bibinfo {author} {\bibfnamefont {N.}~\bibnamefont
  {Spaldin}},\ }\href@noop {} {\bibfield  {journal} {\bibinfo  {journal} {Phys.
  Rev. B}\ }\textbf {\bibinfo {volume} {82}},\ \bibinfo {pages} {220402}
  (\bibinfo {year} {2010})}\BibitemShut {NoStop}%
\bibitem [{\citenamefont {Liu}\ \emph {et~al.}(2020)\citenamefont {Liu},
  \citenamefont {He}, \citenamefont {Kim}, \citenamefont {Khmelevskyi},
  \citenamefont {Toschi}, \citenamefont {Kresse},\ and\ \citenamefont
  {Franchini}}]{liupt2020}%
  \BibitemOpen
  \bibfield  {author} {\bibinfo {author} {\bibfnamefont {P.}~\bibnamefont
  {Liu}}, \bibinfo {author} {\bibfnamefont {J.}~\bibnamefont {He}}, \bibinfo
  {author} {\bibfnamefont {B.}~\bibnamefont {Kim}}, \bibinfo {author}
  {\bibfnamefont {S.}~\bibnamefont {Khmelevskyi}}, \bibinfo {author}
  {\bibfnamefont {A.}~\bibnamefont {Toschi}}, \bibinfo {author} {\bibfnamefont
  {G.}~\bibnamefont {Kresse}},\ and\ \bibinfo {author} {\bibfnamefont
  {C.}~\bibnamefont {Franchini}},\ }\href@noop {} {\bibfield  {journal}
  {\bibinfo  {journal} {Phys. Rev. Mater.}\ }\textbf {\bibinfo {volume} {4}},\
  \bibinfo {pages} {045001} (\bibinfo {year} {2020})}\BibitemShut {NoStop}%
\bibitem [{Note4()}]{Note4}%
  \BibitemOpen
  \bibinfo {note} {The $M_x$ and $M_z$ are contributed by the both the spin and
  orbital magnetic moments.}\BibitemShut {Stop}%
\bibitem [{\citenamefont {Pandey}\ \emph {et~al.}(2017)\citenamefont {Pandey},
  \citenamefont {Mahadevan},\ and\ \citenamefont {Sarma}}]{hubbard2017}%
  \BibitemOpen
  \bibfield  {author} {\bibinfo {author} {\bibfnamefont {S.~K.}\ \bibnamefont
  {Pandey}}, \bibinfo {author} {\bibfnamefont {P.}~\bibnamefont {Mahadevan}},\
  and\ \bibinfo {author} {\bibfnamefont {D.~D.}\ \bibnamefont {Sarma}},\
  }\href@noop {} {\bibfield  {journal} {\bibinfo  {journal} {EPL}\ }\textbf
  {\bibinfo {volume} {117}},\ \bibinfo {pages} {57003} (\bibinfo {year}
  {2017})}\BibitemShut {NoStop}%
\bibitem [{\citenamefont {Middey}\ \emph {et~al.}(2014)\citenamefont {Middey},
  \citenamefont {Debnath}, \citenamefont {Mahadevan},\ and\ \citenamefont
  {Sarma}}]{naoso3model2}%
  \BibitemOpen
  \bibfield  {author} {\bibinfo {author} {\bibfnamefont {S.}~\bibnamefont
  {Middey}}, \bibinfo {author} {\bibfnamefont {S.}~\bibnamefont {Debnath}},
  \bibinfo {author} {\bibfnamefont {P.}~\bibnamefont {Mahadevan}},\ and\
  \bibinfo {author} {\bibfnamefont {D.~D.}\ \bibnamefont {Sarma}},\ }\href@noop
  {} {\bibfield  {journal} {\bibinfo  {journal} {Phys. Rev. B}\ }\textbf
  {\bibinfo {volume} {89}},\ \bibinfo {pages} {134416} (\bibinfo {year}
  {2014})}\BibitemShut {NoStop}%
\bibitem [{\citenamefont {Ueno}\ \emph {et~al.}(2011)\citenamefont {Ueno},
  \citenamefont {Nakamura}, \citenamefont {Shimotani}, \citenamefont {Yuan},
  \citenamefont {Kimura}, \citenamefont {Nojima}, \citenamefont {Aoki},
  \citenamefont {Iwasa},\ and\ \citenamefont {Kawasaki}}]{doping1}%
  \BibitemOpen
  \bibfield  {author} {\bibinfo {author} {\bibfnamefont {K.}~\bibnamefont
  {Ueno}}, \bibinfo {author} {\bibfnamefont {S.}~\bibnamefont {Nakamura}},
  \bibinfo {author} {\bibfnamefont {H.}~\bibnamefont {Shimotani}}, \bibinfo
  {author} {\bibfnamefont {H.~T.}\ \bibnamefont {Yuan}}, \bibinfo {author}
  {\bibfnamefont {N.}~\bibnamefont {Kimura}}, \bibinfo {author} {\bibfnamefont
  {T.}~\bibnamefont {Nojima}}, \bibinfo {author} {\bibfnamefont
  {H.}~\bibnamefont {Aoki}}, \bibinfo {author} {\bibfnamefont {Y.}~\bibnamefont
  {Iwasa}},\ and\ \bibinfo {author} {\bibfnamefont {M.}~\bibnamefont
  {Kawasaki}},\ }\href@noop {} {\bibfield  {journal} {\bibinfo  {journal} {Nat.
  Nanotechnol.}\ }\textbf {\bibinfo {volume} {6}},\ \bibinfo {pages} {408}
  (\bibinfo {year} {2011})}\BibitemShut {NoStop}%
\bibitem [{\citenamefont {Ma}\ \emph {et~al.}(2021)\citenamefont {Ma},
  \citenamefont {Yang},\ and\ \citenamefont {Chen}}]{doping2}%
  \BibitemOpen
  \bibfield  {author} {\bibinfo {author} {\bibfnamefont {J.}~\bibnamefont
  {Ma}}, \bibinfo {author} {\bibfnamefont {R.}~\bibnamefont {Yang}},\ and\
  \bibinfo {author} {\bibfnamefont {H.}~\bibnamefont {Chen}},\ }\href@noop {}
  {\bibfield  {journal} {\bibinfo  {journal} {Nat. Commun.}\ }\textbf {\bibinfo
  {volume} {12}},\ \bibinfo {pages} {2314} (\bibinfo {year}
  {2021})}\BibitemShut {NoStop}%
\bibitem [{\citenamefont {Dobrovits}\ \emph {et~al.}(2019)\citenamefont
  {Dobrovits}, \citenamefont {Kim}, \citenamefont {Reticcioli}, \citenamefont
  {Toschi}, \citenamefont {Khmelevskyi},\ and\ \citenamefont
  {Franchini}}]{naoso3dop}%
  \BibitemOpen
  \bibfield  {author} {\bibinfo {author} {\bibfnamefont {S.}~\bibnamefont
  {Dobrovits}}, \bibinfo {author} {\bibfnamefont {B.}~\bibnamefont {Kim}},
  \bibinfo {author} {\bibfnamefont {M.}~\bibnamefont {Reticcioli}}, \bibinfo
  {author} {\bibfnamefont {A.}~\bibnamefont {Toschi}}, \bibinfo {author}
  {\bibfnamefont {S.}~\bibnamefont {Khmelevskyi}},\ and\ \bibinfo {author}
  {\bibfnamefont {C.}~\bibnamefont {Franchini}},\ }\href@noop {} {\bibfield
  {journal} {\bibinfo  {journal} {J. Phys.: Condens. Matter}\ }\textbf
  {\bibinfo {volume} {31}},\ \bibinfo {pages} {244002} (\bibinfo {year}
  {2019})}\BibitemShut {NoStop}%
\end{thebibliography}
%

\end{document}